\documentclass[referee,pdflatex,sn-mathphys-num]{sn-jnl}% Math and Physical Sciences Numbered Reference Style 
\usepackage{amsmath,amssymb,amsfonts}%
\usepackage{textcomp}%
\usepackage{manyfoot}%
\usepackage{graphicx}
\usepackage{braket}
\usepackage{hyperref}
\usepackage{multirow}%
\usepackage{multicol}
\usepackage{wrapfig}
\usepackage{newpxtext,newpxmath}
\usepackage{setspace}
\usepackage{subcaption}
\usepackage{xcolor}
\usepackage{placeins}
\usepackage{booktabs}
\usepackage{makecell}
\usepackage{tikz}
\usepackage{adjustbox}
\usepackage{quantikz}
\usepackage{bm}

\newcommand{\vect}[1]{\mathbf{#1}}
\newcommand{\gvect}[1]{\boldsymbol{#1}}

\begin{document}

\title[Quantum Latent Diffusion Models]{Quantum Latent Diffusion Models}

%%=============================================================%%
%% GivenName	-> \fnm{Joergen W.}
%% Particle	-> \spfx{van der} -> surname prefix
%% FamilyName	-> \sur{Ploeg}
%% Suffix	-> \sfx{IV}
%% \author*[1,2]{\fnm{Joergen W.} \spfx{van der} \sur{Ploeg} 
%%  \sfx{IV}}\email{iauthor@gmail.com}
%%=============================================================%%

\author[1]{\fnm{Francesca} \sur{De~Falco}}\email{francesca.defalco@uniroma1.it}

\author*[1]{\fnm{Andrea} \sur{Ceschini}}\email{andrea.ceschini@uniroma1.it}
%\equalcont{These authors contributed equally to this work.}

\author[2]{\fnm{Alessandro} \sur{Sebastianelli}}\email{alessandro.sebastianelli@esa.int}
%\equalcont{These authors contributed equally to this work.}

\author[2]{\fnm{Bertrand} \sur{Le~Saux}}\email{bls@ieee.org}

\author*[1]{\fnm{Massimo} \sur{Panella}}\email{massimo.panella@uniroma1.it}

\affil*[1]{\orgdiv{Department of Information Engineering, Electronics and Telecommunications}, \orgname{University of Rome ``La Sapienza''}, \orgaddress{\street{Via Eudossiana 18}, \city{Rome}, \postcode{00184}, \country{Italy}}}

\affil[2]{\orgdiv{$\Phi$-lab, ESRIN}, \orgname{European Space Agency (ESA)}, \orgaddress{\street{Via Galileo Galilei, 1}, \city{Frascati}, \postcode{00044}, \country{Italy}}}

%%==================================%%
%% Sample for unstructured abstract %%
%%==================================%%

\abstract{The introduction of quantum concepts is increasingly making its way into generative machine learning models. However, while there are various implementations of quantum Generative Adversarial Networks, the integration of quantum elements into diffusion models remains an open and challenging task. In this work, we propose a potential version of a quantum diffusion model that leverages the established idea of classical latent diffusion models. This involves using a traditional autoencoder to reduce images, followed by operations with variational circuits in the latent space. To effectively assess the benefits brought by quantum computing, the images generated by the quantum latent diffusion model have been compared to those generated by a classical model with a similar number of parameters, evaluated in terms of quantitative metrics. The results demonstrate an advantage in using a quantum version, as evidenced by obtaining better metrics for the images generated by the quantum version compared to those obtained by the classical version. Furthermore, quantum models continue to outperform even when considering small percentages of the dataset for training, demonstrating the quantum's ability to extract features more effectively even in a few shot learning scenario.}


%%\pacs[JEL Classification]{D8, H51}

%%\pacs[MSC Classification]{35A01, 65L10, 65L12, 65L20, 65L70}

\maketitle

\section{Introduction}
Within the framework of classical Artificial Intelligence (AI), Diffusion Models (DMs) have emerged as top contenders for generating data and images, as evidenced by  \cite{pmlr-v37-sohl-dickstein15, ho2020denoising, Rombach2021HighResolutionIS}. DMs employ an iterative diffusion process, progressively enhancing the representation of complex distributions by refining the data distribution through a sequence of diffusion steps.
They exhibit superior quality and training stability compared to state-of-the-art Generative Adversarial Networks (GANs) \cite{NIPS2014_5ca3e9b1}. 
In particular, latent DMs \cite{Rombach2021HighResolutionIS} have gained prominence as they enable operation in a computationally more suitable space, transitioning from pixel space to latent space. The advantage is evident, facilitating the efficient processing of even large-scale images. Additionally, there is a substantial improvement in terms of computation time and conservation of energy resources.

Meanwhile, various proposals have been made in the introduction of quantum computing architectures into generative models, as over the years, the search for supervised \cite{farhi2018classification, johri:hal-03432449, Schuld_2017} and unsupervised \cite{aimeur:hal-00736948, Benedetti_2019} machine learning algorithms has highlighted how quantum can bring various benefits \cite{Bravyi_2018, Bravyi_2020, Abbas_2021}.
In particular, concerning Quantum Generative Adversarial Networks (QGANs), there have been various implementations \cite{Huang_2021, chang2024latentstylebasedquantumgan, Tsang2022HybridQG, NIPS2017_8a1d6947} that have highlighted how even just making the generator quantum can bring benefits, resulting in the generation of better images compared to those generated by a classical GAN. Moreover, a quantum GAN can achieve better performance than its classical counterpart even when the total number of trainable parameters is fewer in the quantum case. 

Specifically, in \cite{chang2024latentstylebasedquantumgan} a quantum GAN is presented, which is formed solely by the quantum generator while the discriminator remains classic, operating in the latent space. In other words, both the generator and the discriminator work not in the pixel space but in the latent space, which is reached through the use of a pre-trained autoencoder on the referenced dataset. The advantages highlighted by this architecture include superior image quality generated by the quantum GAN, evaluated in terms of predefined metrics such as Fréchet Inception Distance (FID) and Jensen-Shannon Divergence (JSD). Additionally, it is noteworthy that the quantum GAN demonstrates superior performance even when trained for a reduced number of epochs or with a limited portion of the dataset. This underscores how the introduction of quantum elements can enhance the model's ability to extract features even with limited training epochs or a restricted dataset. As for DMs, there have only been recent attempts to explore their quantum adoption \cite{cacioppo2023quantum, parigi2023quantumnoisedriven}.

While a theoretical discussion of a potential quantum generalization of diffusion models is proposed only in \cite{parigi2023quantumnoisedriven}, with results tied only to very basic and simplified scenarios, two different architectures are proposed in \cite{cacioppo2023quantum}: the first one works on downsized images from the MNIST dataset; the second model suggests operating on a latent space using a pre-trained autoencoder.
Our idea, instead, aims at always working in a latent space in order to operate in a computationally more suitable space.
Accordingly, in this paper we propose a quantum latent diffusion model (QLDM), which is characterized by the use of three different variational quantum circuits (VQCs), allowing us to work not only on image-related information but also on temporal information of the diffusion process. Moreover, because we work in the latent space, it is possible to use angle encoding as data encoding, which can be more efficient than amplitude encoding since the latter requires an exponential number of circuit runs to obtain the output distribution.

In order to evaluate the actual efficiency of our QLDM approach, various analyses were conducted always comparing the results obtained from our model with an equivalent classical one. Precisely, from the results obtained on MNIST and Fashion MNIST datasets, we will demonstrate how our model achieves better or comparable performances to the ones of the classical model, with the additional advantage of achieving them with a lower number of epochs and especially with a lower amount of training data compared to the classical counterpart. Additionally, the analysis of the loss obtained during training also demonstrates once again how the QLDM converges with a lower number of iterations compared to the classical case.

The paper is organized as follows. In Sect.~\ref{sec:background}, a background is provided on variational circuits and classical diffusion models; the adopted methodology is explained in Sect.~\ref{sec:methodology}, while the obtained results are discussed in Sect.~\ref{sec:results}. Finally, our conclusions are drawn in Sect.~\ref{sec:conclusions}.

\section{Theoretical Background}
\label{sec:background}
\subsection{Classical diffusion models}
In recent years, DMs have proven to be an important class of generative models. A standard mathematical formulation for diffusion models is the one presented by \cite{ho2020denoising}, which is summarized in the following in order to provide readers with a general overview of its foundations. 

\begin{figure}[!ht]
    \centering
    \includegraphics[width=0.9\columnwidth]{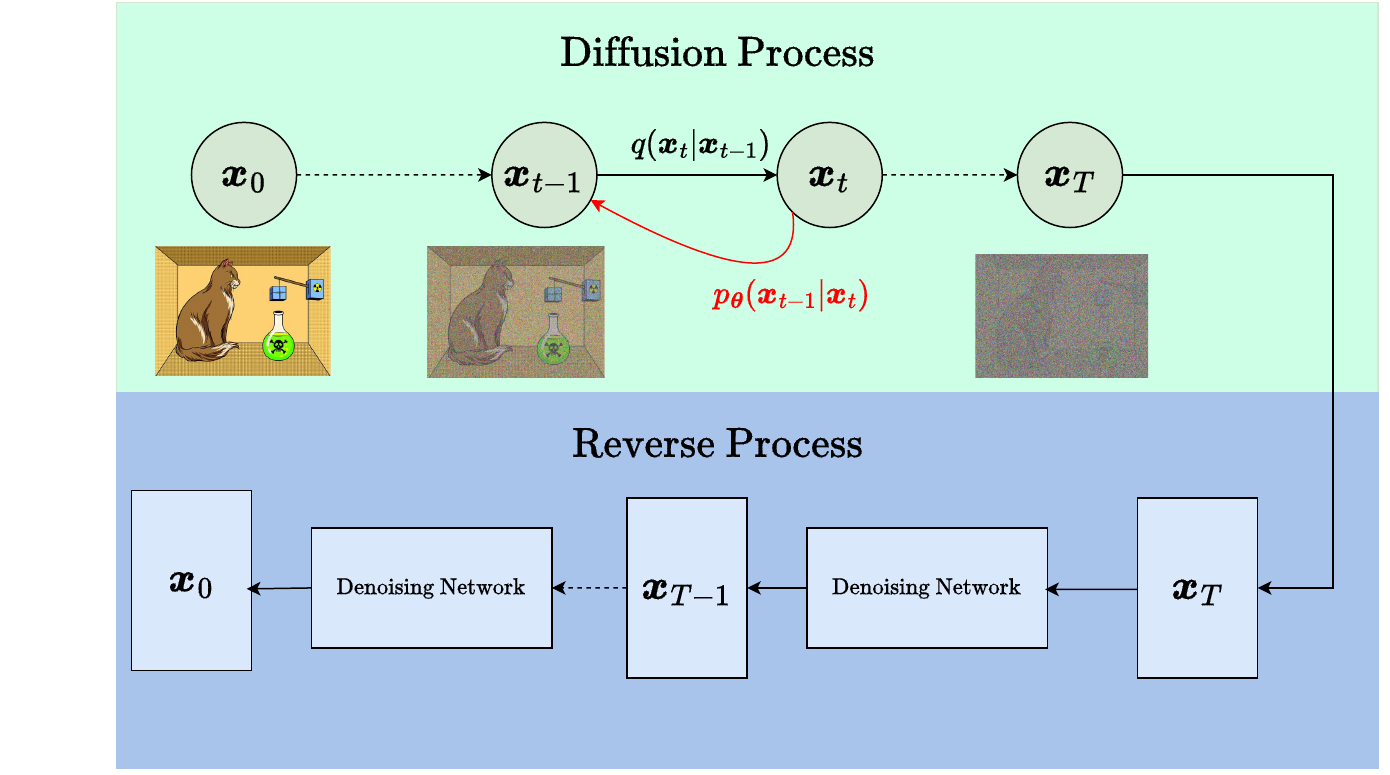}
		\vspace{6pt}
    \caption{\small Diffusion process and reverse process of a DM.}
    \label{fig:diffusion process}
\end{figure}

DMs mainly consist of two distinct phases as shown in Fig.~\ref{fig:diffusion process}. The first one is the forward process, also called diffusion, involving a transformation that gradually converts the original data distribution ${\vect{x}_0\sim q}$, where $q$ is a probability distribution to be learned, by repeatedly adding Gaussian noise:
\begin{equation}
	 \vect{x}_t = \sqrt{1 -\beta_{t}}\vect{x}_{t-1} + \sqrt{\beta_{t}}\gvect{\epsilon}_t,\,\,t=1\dots T\,,
\end{equation}
where ${\gvect{\epsilon}_1,\dots,\gvect{\epsilon}_T}$ are IID samples drawn from a zero-mean, unit variance Gaussian (normal) distribution $\mathcal{N}(\vect{0}, \vect{I})$, and $\beta_{t}$ determines the variance scale for the $t$-th step.
This progression is underpinned by a Markov chain that can be represented as follows:
\begin{align}
	&q(\vect{x}_{0:T}) = q(\vect{x}_0)\prod_{t=1}^{T}q(\vect{x}_{t}|\vect{x}_{t-1})\,,\\
	&q(\vect{x}_{t}|\vect{x}_{t-1})= \mathcal{N}\left(\sqrt{1 -\beta_{t}}\vect{x}_{t-1}, \beta_{t}\vect{I}\right)\,,	
\end{align}
being $\vect{I}$ the identity matrix. 

The goal of the forward process is to add incremental noise to the initial sample $\vect{x}_0$ over a certain number of steps, until at the final time step $T$ all traces of the original distribution ${\vect{x}_0\sim q}$ are lost so as to obtain ${\vect{x}_T\sim \mathcal{N}(\vect{0}, \vect{I})}$.
Through the application of the `reparameterization trick', a closed-form solution becomes available for calculating the total noise at any desired step using the cumulative product:
\begin{equation}
\vect{x}_{t} = \sqrt{\bar\alpha_{t}}\,\vect{x}_{0}+ \sqrt{1 -\bar\alpha_{t}}\,\gvect{\epsilon}\,,
\end{equation}
where $\alpha_{t} = 1 - \beta_{t}$, $\bar\alpha_{t}= \prod_{i=1}^{t} \alpha_{i}$, and $\gvect{\epsilon}\sim\mathcal{N}(\vect{0}, \vect{I})$ is the Gaussian noise.

The second phase is the reverse process or backward diffusion, where the transformations gradually restore the initial noise distribution and reconstruct a noise-free version of the original data. If we could successfully reverse the aforementioned process sampling from ${q(\vect{x}_{t-1}|\vect{x}_t)}$, we would gain the ability to recreate the true sample starting from the Gaussian noise input ${\vect{x}_T\sim\mathcal{N}(\vect{0}, \vect{I})}$; it is also noteworthy that when $\beta_t$ is sufficiently small, ${q(\vect{x}_{t-1}|\vect{x}_t)}$ is close to a Gaussian distribution. Regrettably, estimating ${q(\vect{x}_{t-1}|\vect{x}_t)}$ is complex due to its reliance on the entire dataset and hence, a data-driven learning model like a neural network must be used in order to approximate these conditional probabilities, enabling the execution of the reverse diffusion process.

Let $p_{\gvect{\theta}}$ be the mathematical model depending on some parameters $\gvect{\theta}$ that represents the estimated distribution of the backward diffusion process:
\begin{align}
	&p_{\gvect{\theta}}(\vect{x}_T) = \mathcal{N}(\vect{0}, \vect{I})\,,\\
	&p_{\gvect{\theta}}(\vect{x}_{t-1}|\vect{x}_{t})= \mathcal{N}\left(\gvect{\mu}_{\gvect{\theta}}(\vect{x}_{t},t), \gvect{\Sigma}_{\gvect{\theta}}(\vect{x}_{t},t)\right)\,,	
\end{align}
where $\gvect{\mu}_{\gvect{\theta}}(\vect{x}_{t},t)$ and $\gvect{\Sigma}_{\gvect{\theta}}(\vect{x}_{t},t)$ are the general outputs of the adopted neural network, which takes as inputs $\vect{x}_{t}$ and $t$.

A simplified approach based on variational inference assumes a fixed covariance matrix, such as for instance ${\gvect{\Sigma}_{\gvect{\theta}}(\vect{x}_{t},t)=\beta_t\vect{I}}$, and the direct estimation by the neural network of the noise $\gvect{\epsilon}_{\gvect{\theta}}(\vect{x}_{t},t)$ at time step $t$. Then, using reparameterization and the normal distributions of conditional data, we obtain:
\begin{equation}
	\gvect{\mu}_{\gvect{\theta}}(\vect{x}_{t},t) = {\frac{1}{\sqrt{\alpha_t}}} \left(\vect{x}_t - \frac{1-\alpha_t}{\sqrt{1-\bar\alpha_t}}\gvect{\epsilon}_{\gvect{\theta}}(\vect{x}_{t},t)\right)\,.
\end{equation}
The neural network producing $\gvect{\epsilon}_{\gvect{\theta}}(\vect{x}_{t},t)$ is usually trained by stochastic gradient descent on an even more simplified loss function like:
\begin{equation}
L^\mathrm{simple}_t = \mathbb{E}_{\vect{x}_0\sim q,t,\gvect{\epsilon}\sim\mathcal{N}(\vect{0},\vect{I})} \left[ \left\| \gvect{\epsilon}_{\gvect{\theta}}(\vect{x}_{t},t) - \gvect{\epsilon} \right\|^2 \right]\,.
\end{equation}

%\FloatBarrier

\subsection{Variational Quantum Circuits}
 Variational Quantum Algorithms (VQAs) are the most common quantum machine learning (QML) algorithm currently implemented on today's quantum computers \cite{Cerezo_2021}. Based on the several ways they can be adopted, using the most common optimized versions of machine learning approaches \cite{scala_2023,Incudini2023}, VQAs make use of parametrized quantum circuits known as ansatzes. Ansatz circuits are composed of quantum gates that manipulate qubits through specific parametrized unitary operations. However, these operations depend on parameters denoted as $\gvect{\theta}$, which are the parameters to be trained during the training process.
 
\begin{figure}[!ht]
    \centering
    \includegraphics[trim={0 0 2.5cm 0}, width=0.85\columnwidth]{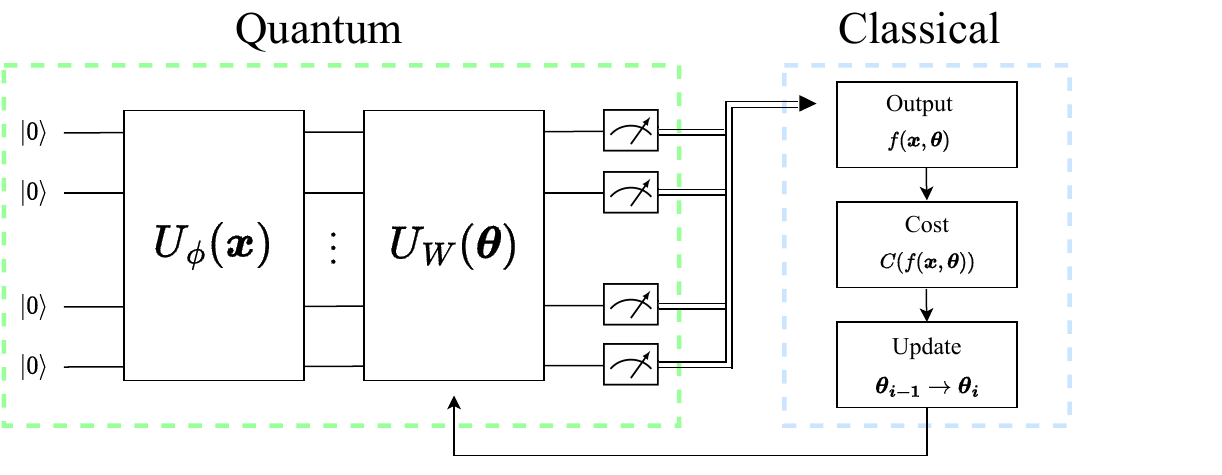}
		\vspace{6pt}
    \caption{\small Scheme of a hybrid quantum-classical VQC.}
    \label{vqc}
\end{figure}

The training workflow of a VQC, shown in Fig.~\ref{vqc}, can be summarized as follows:
\begin{enumerate}
	\item classical data $\gvect{x} \in \mathbb{R}^{n}$ is appropriately encoded in a quantum state of the Hilbert space $\mathbb{H}^{2^{n}}$ through the unitary $U_{\phi}({\gvect{x}})$, to be used by the quantum computer;
	\item an ansatz $U_W({\gvect{\theta}})$ of $\gvect{\theta}$-parametrized unitaries with randomly initialized parameters and fixed entangling gates is applied to the quantum state $|\phi({\gvect{x}})\rangle$ obtained after the encoding;
	\item upon completion, measurements are taken to obtain the desired outcomes. The expected value with respect to a given observable $\hat{O}$ is typically computed, and the resulting prediction is given by: 
			\begin{equation}
			f({\gvect{x}}, \gvect{\theta}) = \langle \phi(\gvect{x}) | U_W(\gvect{\theta})^\dagger \hat{O} U_W(\gvect{\theta}) | \phi(\gvect{x}) \rangle\,;
			\end{equation}
    \item finally, a suitable loss function is evaluated, and a classical co-processor is used to properly update the parameters $\gvect{\theta}$. 
\end{enumerate}
This cycle is repeated until a termination condition is met. To update $\gvect{\theta}$ and train the VQC, gradient-based techniques can be used; gradients in a parametrized quantum circuit are calculated via the parameter-shift rule:
\begin{equation}
     \nabla_{\theta}f(\gvect{x},\theta) = \frac{1}{2} \left[ f(\gvect{x},\theta + \frac{\pi}{2}) - f(\gvect{x},\theta - \frac{\pi}{2}) \right]\,,
\end{equation}
where $f(\gvect{x}, \theta)$ is the output of the quantum circuit and $\theta$ is the parameter to be optimized. 

\section{Proposed Methodology}
\label{sec:methodology}

As discussed in Sect.~\ref{sec:background}, the classic diffusion model involves the use of a Markov chain in the forward phase, which introduces noise through a Markov kernel. A parametric model is then trained so that in the reverse phase, it can reproduce the inverse Markov chain. This classic diffusion process remains unchanged in our work, with the only difference being that the parametric model used is composed of VQCs.

\begin{figure}[!ht]
    \centering
    \includegraphics[width=\columnwidth]{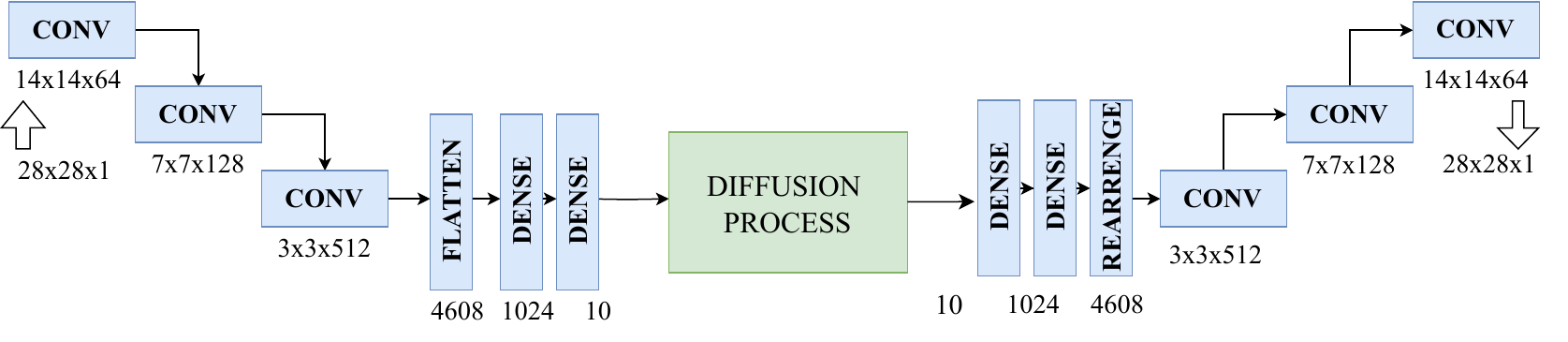}
		%\vspace{3pt}
    \caption{\small Architecture of quantum latent diffusion model and classical convolutional autoencoder.}
    \label{quantum latent diffusion}
\end{figure}

In order to efficiently introduce variational quantum circuits while maintaining a limited number of qubits, we chose to work not in the pixel space but in a latent space. Our architecture, therefore, as depicted in Fig.~\ref{quantum latent diffusion}, initially consists of a classical convolutional autoencoder trained separately on the dataset of interest. The presence of the autoencoder not only allows us to transition from the pixel space to vectors of dimension 10 but also introduces strong non-linearity, benefiting our QLDM.

Our QLDM operates similarly to classical Latent Diffusion Models (LDMs) in terms of its overall structure and function. The primary difference lies in the utilization of VQCs instead of classical neural networks, such as Multi-Layer Perceptrons (MLPs), for the denoising process. Specifically, the diffusion process is classical while the denoising phase is implemented through three distinct VQCs, each operating under quantum dynamics.

\begin{figure}[!ht]
    \centering
    \includegraphics[width=0.9\columnwidth]{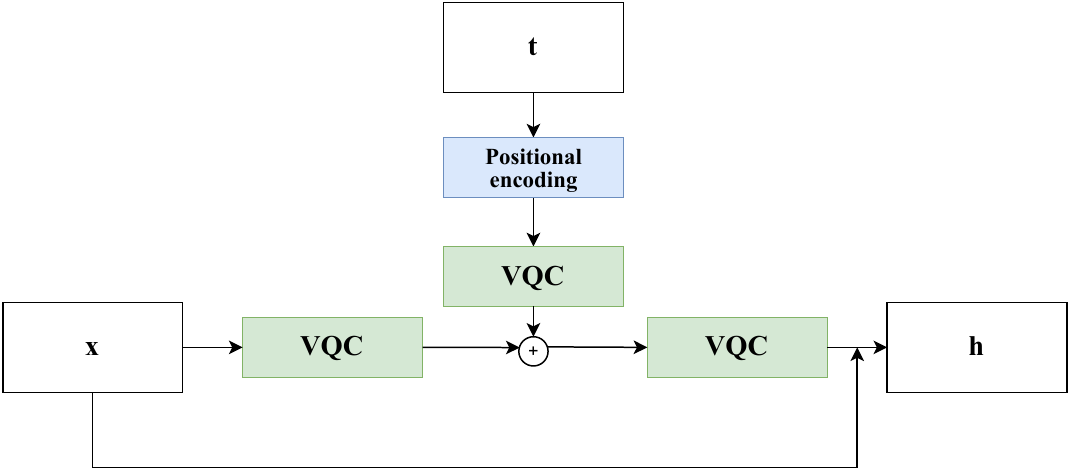}
		\vspace{6pt}
    \caption{\small Architecture of the proposed quantum latent diffusion model. It consists of three different variational circuits: the first processes only the input latent vector; the second works on the temporally embedded sinusoidal information; the third works on both the latent vector of the image and the temporal information.}
    \label{quantum latent diffusion1}
\end{figure}

The architecture of the QLDM is therefore shown in Fig.~\ref{quantum latent diffusion1}: it essentially consists of three different VQCs that play different roles and their arrangement seeks to mimic the classical ResNet model \cite{He_2016_CVPR}. The first of the three VQCs processes the latent vector $\vect{x}$, which contains information about the image previously downscaled by the encoder and degraded through the forward process. The second VQC processes the temporal information, previously encoded classically through positional encoding \cite{vaswani2023attention}. The inclusion of this VQC is crucial to incorporate the diffusion times into the model. Finally, the third VQC is aimed at working on both temporal and image-related information, taking as input the sum of the outputs of the two previous VQCs. Additionally, there is a skip connection, that adds $\vect{x}$ to the output of the third VQC, as in classical ResNet, aiming to enable faster convergence and enhance learning accuracy. 

The expectation values measured from the third VQC and combined with the skip connection are used for the direct estimation of the noise $\gvect{\epsilon}_{\gvect{\theta}}(\vect{x}_{t},t)$ at time step $t$. The advantages of employing a quantum denoising process over a classical one include the potential for leveraging quantum mechanical properties such as superposition and entanglement. These properties can enhance the efficiency of generating high-dimensional data (e.g., images), thereby speeding up data processing tasks \cite{lloyd2014quantum,wiebe2012quantum,yao2017quantum}. Additionally, existing literature provides evidence of quantum advantages when using VQCs over Deep Neural Networks (DNNs). For example, VQCs have been shown to outperform classical DNNs in generative tasks \cite{du2020expressive} and demonstrate an exponential advantage in model size for function approximation of high-dimensional smooth functions \cite{yu2023provable}. Hence, it is plausible that employing VQCs instead of MLPs in our context can yield similar benefits.

Regarding the structure of each variational circuit, the use of angle encoding has been preferred over amplitude encoding, which is less efficient. This is because amplitude encoding requires multiple circuit runs, and the number of runs grows exponentially as the dimension of the data to be mapped increases.
Angle encoding is possible and used effectively because the latent vector to be mapped into a quantum state is of small dimensions, specifically 10, corresponding to the number of qubits to be used.

\begin{figure}[!ht]
    \centering
    \includegraphics[width=0.55\linewidth]{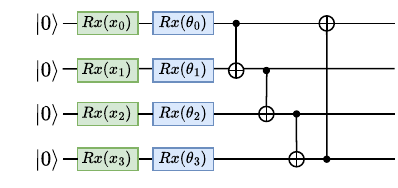}
		\vspace{6pt}
    \caption{\small The basic ansatz used, characterized by rotations only on RX, preceded by angle encoding.}
    \label{fig: basic ansatz}
\end{figure}

\begin{figure}[!ht]
    \centering
    \includegraphics[width=0.9\linewidth]{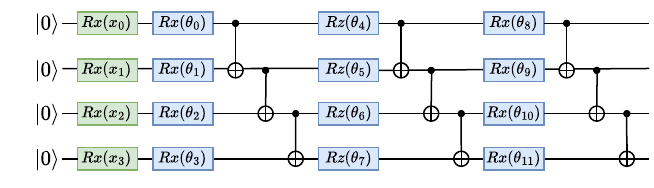}
		\vspace{6pt}
    \caption{\small The more expressive ansatz used, characterized by rotations on RX, followed by RZ and finally RX, preceded by angle encoding.}
    \label{fig:rxrzrx}
\end{figure}

As for the ansatz, two different ones have been employed: the first, the simpler one, is the basic one shown in Fig.~\ref{fig: basic ansatz}, composed only by RX rotations. The second circuit used is a more expressive one in Fig.~\ref{fig:rxrzrx}, composed of a rotation on RX, followed by RZ, and finally RX. This way, it effectively covers the entire Bloch sphere, providing greater expressiveness to the circuit. The measurement was performed either on the observable Z or X, allowing us to evaluate the impact that the use of either measurement can have on the model, as already proposed in \cite{zaman2024studying}. Additionally, the depth of each ansatz was fixed at either 3 or 4.  

To always have a comparison with a classical model, to observe the quantum behavior consistently, a classical network has been implemented that perfectly replicates the fully quantum architecture and is implemented with simple fully connected layers with an input of 10 and an output of 10. Finally, paying attention to the total number of trainable parameters, in the case of the classical model, there are 330 parameters, whereas in the case of the QLDM constructed with the basic ansatz, there are 120 parameters. Meanwhile, if the ansatz used involves rotations on RX-RZ-RZ, there are 270 parameters if the depth is 3, otherwise 360 if the depth is 4. The number of parameters between the classical and quantum models are therefore comparable.

%\FloatBarrier

%\end{multicols}

%\begin{multicols}{2}
    
%\FloatBarrier
%\hfill
\FloatBarrier
\hfill

\section{Experimental Results}
In this section, we analyze the results obtained by evaluating the images generated by our architecture on the MNIST and Fashion MNIST datasets and comparing them with those obtained by the classical architecture.
\label{sec:results}
\subsection{Experimental settings}
The implementation is carried out in Python 3.8 using PennyLane \cite{bergholm2022pennylane} and Flax \cite{flax2020github}. PennyLane is a framework which enables local quantum circuits simulations and integration with classical neural networks, whereas Flax is an open-source machine learning framework that provides a flexible and efficient platform for hybrid neural network execution via compilation. 
We use PennyLane for the implementation of quantum circuits, while the classical and quantum networks and the entire training process are carried out in Flax. 

Regarding the experiments, we use the following benchmark datasets: MNIST \cite{lecun2010mnist}; Fashion MNIST \cite{DBLP:journals/corr/abs-1708-07747}; EuroSAT \cite{helber2017eurosat}. MNIST and Fashion MNIST contain grayscale images belonging to 10 different classes, with a total of 60k training samples. EuroSAT is instead a dataset based on Sentinel-2 satellite images, consisting of 10 classes with 270,000 labeled and geo-referenced samples. We used the version containing only the optical R, G, B frequency bands encoded as JPEG images, and specifically focused on only two classes: Forest (class number 1) and HerbaceousVegetation (class number 2), with images resized to 28x28 dimensions .
The data from all three dataset are initially scaled between -1 and 1.

The L2 loss is used with the P2 weighting \cite{choi2022perception} for training both the quantum diffusion models and the classical one. We use an exponential moving average (EMA) over model parameters with a rate that depends on the training step, and the Adam optimizer \cite{kingma2014adam} is used with a learning rate of $10^{-3}$, $\beta_1$ of 0.9, and $\beta_2$ of 0.99. The training process consists of a total of 40 epochs. Such hyperparameters were chosen after an extensive grid search procedure.

The metrics used for evaluations are FID \cite{NIPS2017_8a1d6947}, Kernel Inception Distance (KID) \cite{NIPS2016_8a3363ab}, and Inception Score (IS) \cite{betzalel2022study}, assessed on 10000 generated images for MNIST and Fashion MNIST, while 5400 for EuroSAT, as to have a number equal to the real images of the two classes in the dataset. We utilize the TorchMetrics library \cite{Detlefsen2022}, replicating  each channel of the generated grayscale images three times to make the dimensions compatible with those required by InceptionV3 network backbone. For the KID calculation, the subset size for computing mean and variance was set to 100, while for the IS calculation, the dataset was divided into 10 splits for mean and variance computation.
A machine equipped with an AMD Ryzen {7\textsuperscript\texttrademark} 5800X 8-Core CPU at 3.80 GHz and with 64 GB of RAM is used for the experiments.

\subsection{MNIST dataset results}
Initially, let us consider the generated images from the MNIST dataset. The architectures under consideration are: 
\begin{itemize}
\item `BasicQ', where the QLDM is realized with the basic ansatz and with measurement performed on observable Z; 
\item `3zQ', realized with the most expressive ansatz visible in Fig.~\ref{fig:rxrzrx} and with an ansatz depth of 3 and measurement performed on observable Z; 
\item `3xQ', still with the ansatz present in Fig.~\ref{fig:rxrzrx} and with measurement performed on observable X; 
\item `4zQ' where now the depth is 4 and the measurement is performed on observable Z;
\item `4xQ', where the depth remains 4 but the measurement is performed on observable X. 
\end{itemize}
The `Basic' architecture performing measurement on observable X was not considered as it showed extremely worse metric values compared to the other quantum architectures considered. All these different quantum models are always compared with a classical model.

The images generated by all models, both classical and quantum, were quantitatively evaluated in terms of commonly used metrics for assessing the quality of generative models, namely FID, KID, and IS. These metrics allow us to understand the similarity between the distributions of produced and real pictures, thus enabling the evaluation of the model's ability to capture the essence of real images.

\begin{figure}[!ht]
  \begin{subfigure}{0.5\textwidth}
    \includegraphics[width=\linewidth]{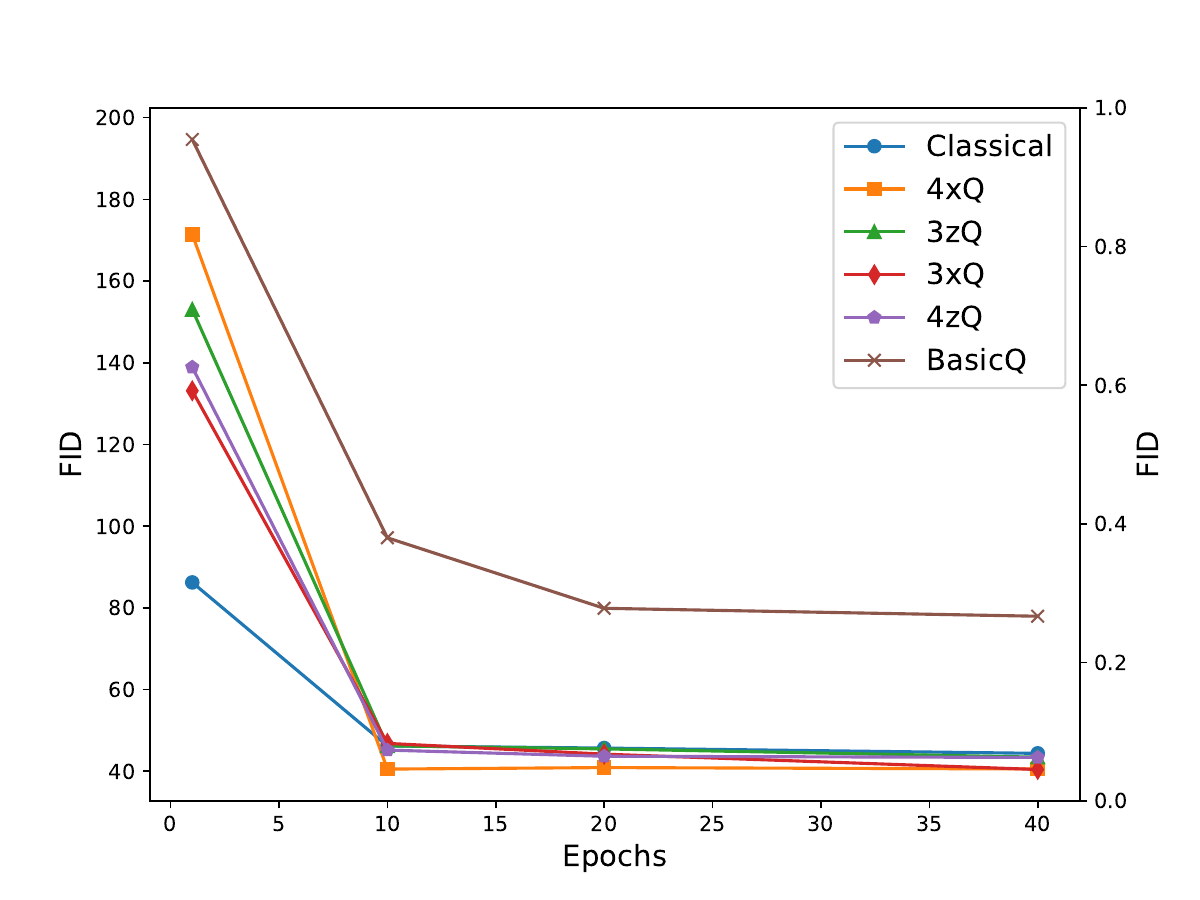}
     \caption{\small }
    \label{fig: fidmnist}
  \end{subfigure}
  \hfill
  \begin{subfigure}{0.5\textwidth}
    \includegraphics[width=\linewidth]{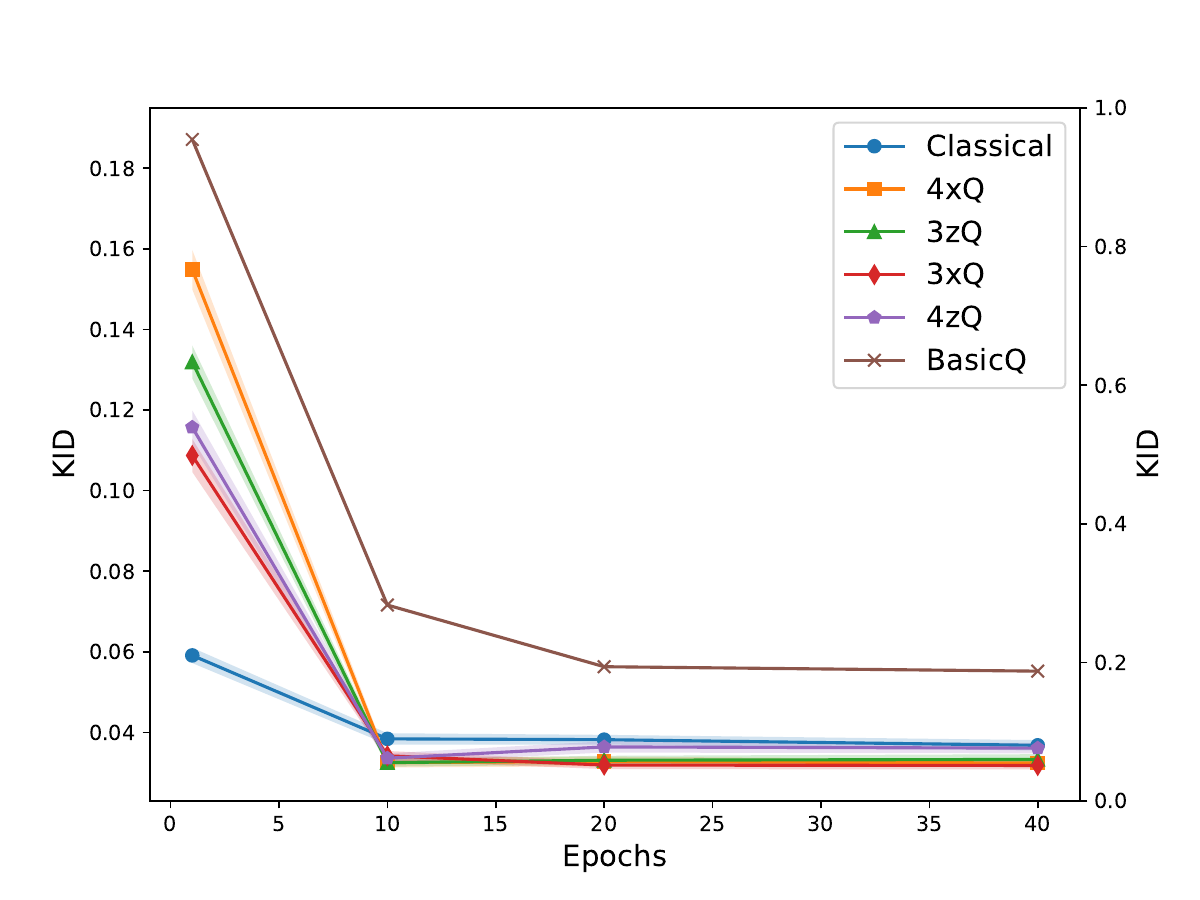}
    \caption{\small }
    \label{fig: kidmnist}
  \end{subfigure}
  \hfill
  \begin{subfigure}{0.5\textwidth}
    \includegraphics[width=\linewidth]{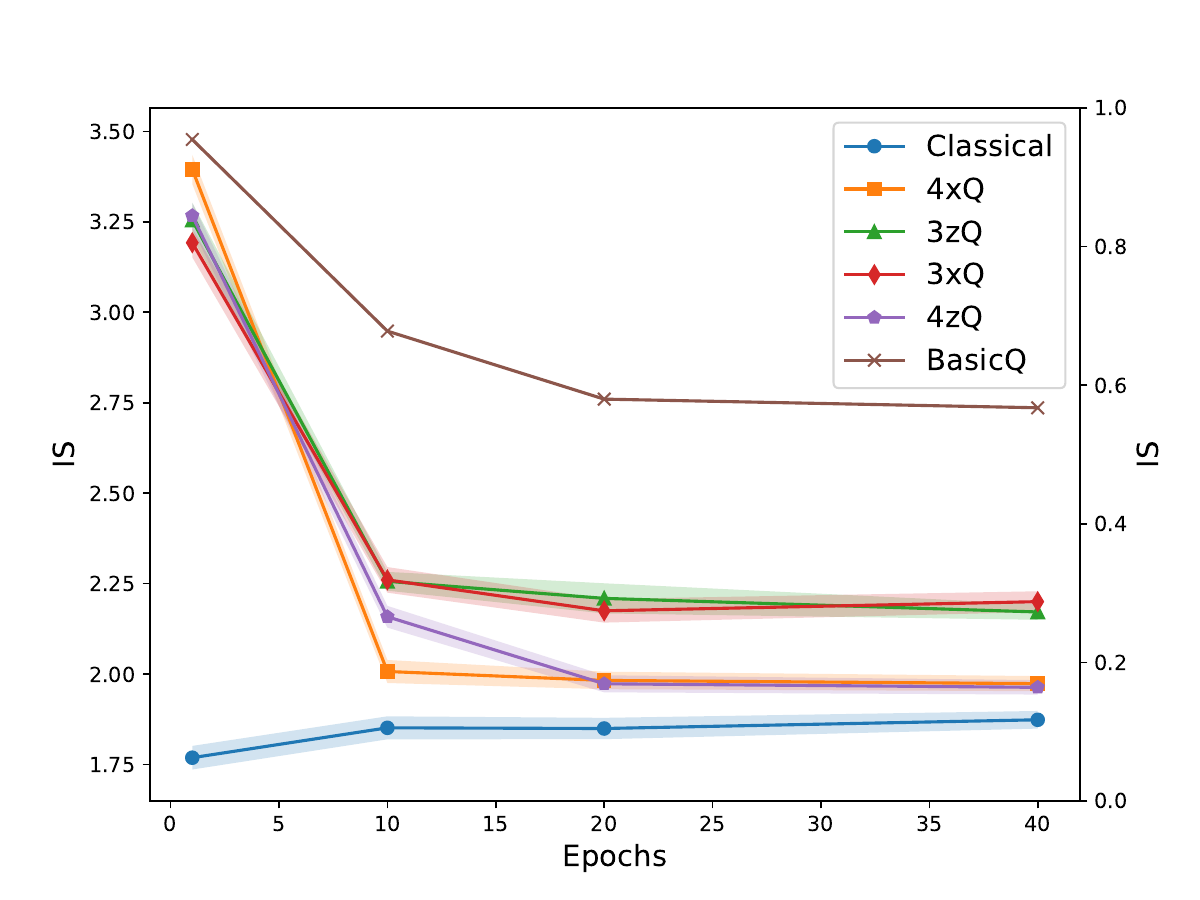}
    \caption{\small }
    \label{fig:ismnist}
    
  \end{subfigure}
    \hfill
  \begin{subfigure}{0.5\textwidth}
    \includegraphics[width=\linewidth]{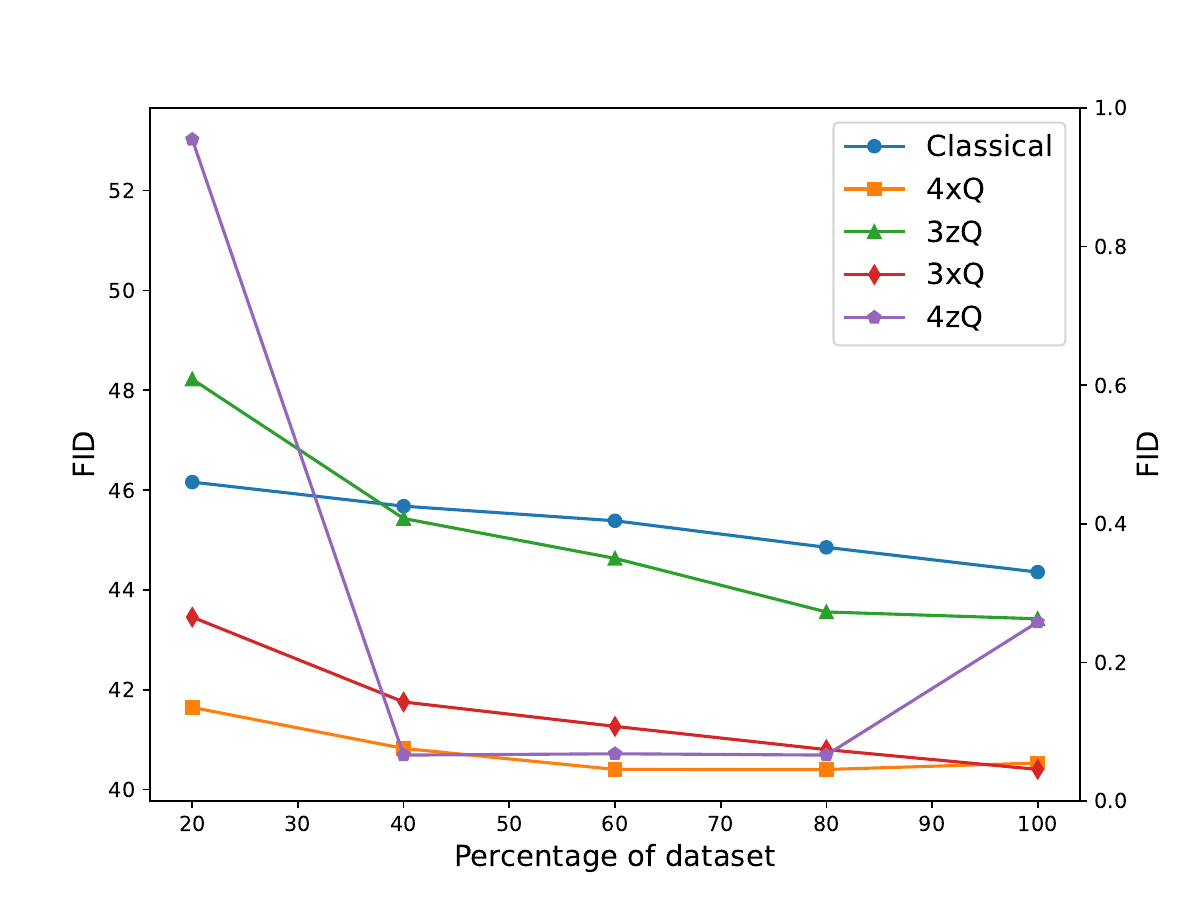}
    \caption{\small }
    \label{fig:fidmnistp}
    
  \end{subfigure}
      \hfill
  \begin{subfigure}{0.5\textwidth}
    \includegraphics[width=\linewidth]{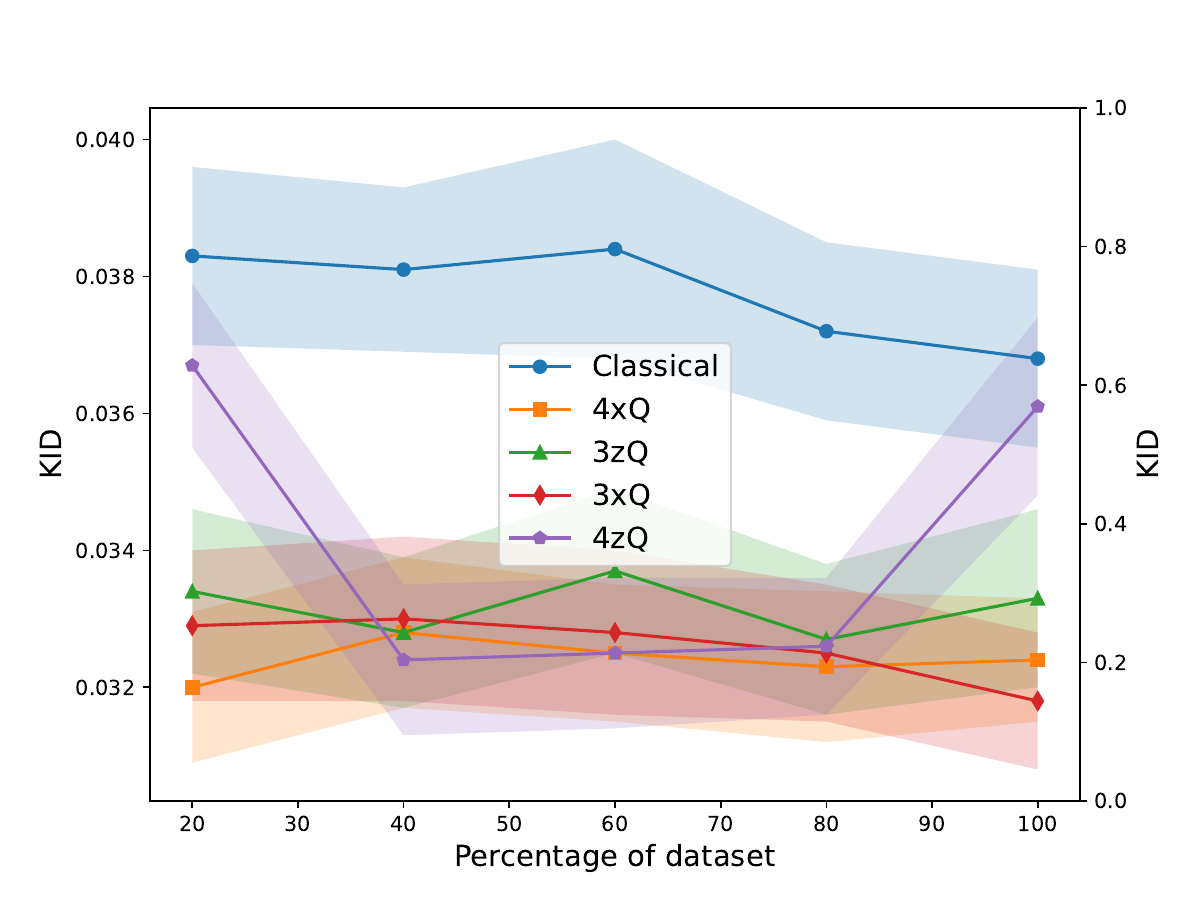}
    \caption{\small }
    \label{fig:kidmnistp}
    
  \end{subfigure}
      \hfill
  \begin{subfigure}{0.5\textwidth}
    \includegraphics[width=\linewidth]{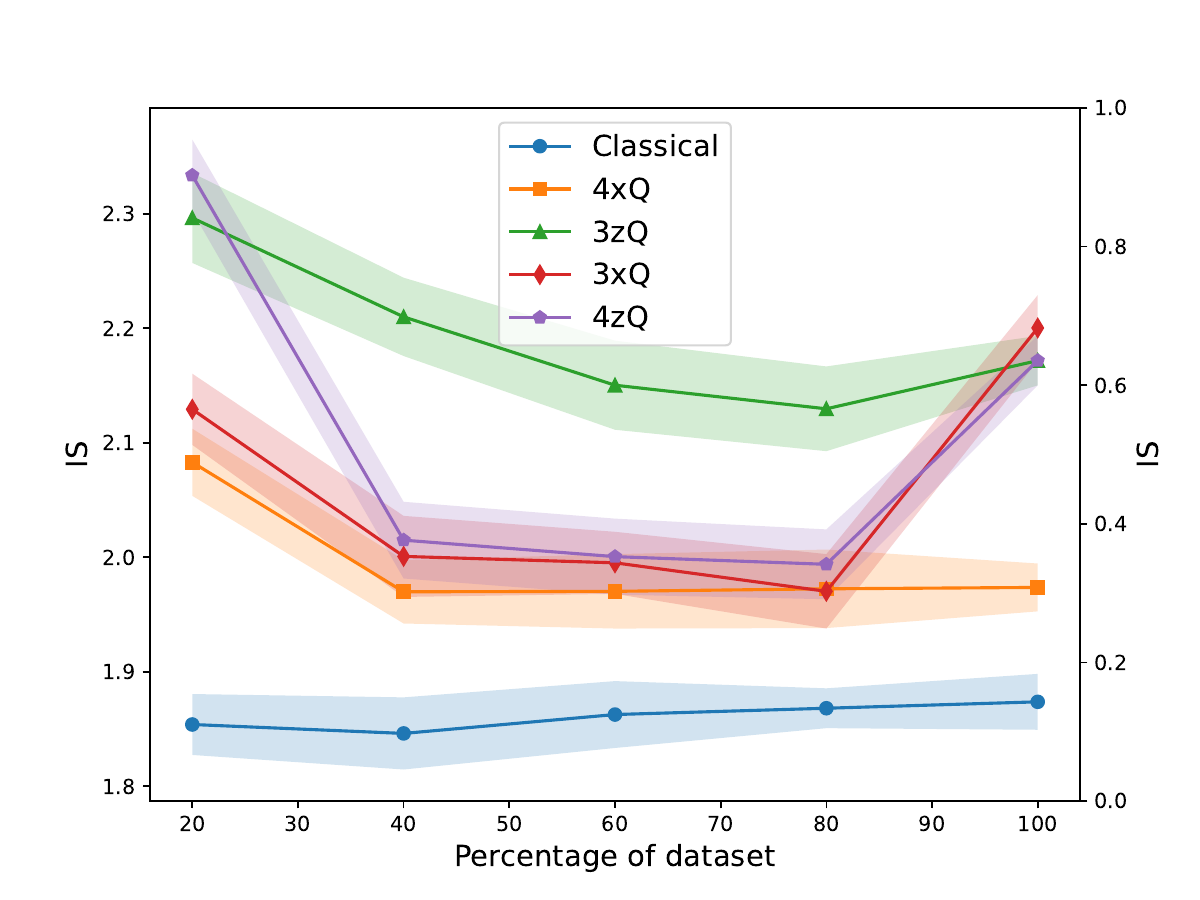}
    \caption{\small }
    \label{fig:ismnistp}
  \end{subfigure}
  \caption{\small Plots MNIST: a) FID plot as a function of training epochs using 100\% of the dataset for model training; b) KID plot as a function of training epochs using 100\% of the dataset for model training; c) IS plot as a function of training epochs using 100\% of the dataset for model training; d) FID plot as a function of dataset percentages; e) KID plot as a function of dataset percentages; f) IS plot as a function of dataset percentages. In the KID and IS plots, the mean and standard deviation are plotted. For the KID, 100 subsets of size 100 were used for computing mean and variance, while for IS, 10 subsets of size 1000 were used.}
  \label{fig: mnist}
\end{figure}

\begin{figure}[!ht]
  \begin{subfigure}{0.4\textwidth}
    \includegraphics[width=\linewidth]{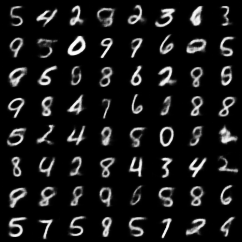}
    \caption{\small }
    \label{fig: mnist classica}
  \end{subfigure}
  \hfill
  \begin{subfigure}{0.4\textwidth}
    \includegraphics[width=\linewidth]{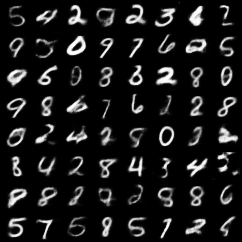}
    \caption{\small }
    \label{fig: mnist 3x}  
  \end{subfigure}
  \caption{\small Generated images of MNIST: a) generated by the Classical Architecture; b) generated by the 3xQ architecture.}
  \label{fig: mnist im}
\end{figure}

The values of FID calculated on the generated images after the model has been trained for a fixed number of epochs. The trend of FID is visible in Fig.~\ref{fig: fidmnist}. As can be seen, the BasicQ architecture, which has only 120 parameters, always presents much higher FID values and therefore worse compared to the other architectures, both quantum and classical. This is attributable to the extremely low number of trainable parameters of the model. The other quantum architectures instead present FID values much higher at the first epoch but progressively decreasing. At the tenth epoch, in fact, the FID values of the images of the quantum architectures become comparable to that of the images of the classical architecture and furthermore, the 4xQ architecture has a significantly better FID already at the tenth epoch. Halfway through training, at the 20th epoch, the quantum models continue to have performance certainly comparable to the classical model and in the case of 4zQ and 4xQ certainly better. At the end of training, i.e. at the 40th epoch, all quantum models except for BasicQ outperform the classical model which reaches an FID of 44.3568, while the best quantum model at the last epoch is 3xQ with an FID of 40.4031 as shown in Table~\ref{tab:tabella1}. The images generated by the classical architecture and the 3xQ architecture are also shown in Fig.~\ref{fig: mnist im}.

\begin{table}[!ht]
  \centering
	\resizebox{\textwidth}{!}{%
  \begin{tabular}{lcccccccc}
    \toprule
     Metrics & Classical& BasicQ & 3zQ & 3xQ & 4zQ & 4xQ & GAN & QGAN\\
    \midrule
    Parameters& 330 & 120& 270 & 270& 360 &360 & 110+231 & 120+231\\
    FID\(\downarrow\) & 44.3568 & 77.9277 & 43.4210 & \textbf{40.4031} & 43.3555& 40.5321 & 56.8930 & 89.9665 \\
    KID\(\downarrow\)& $0.0368 \pm 0.0013$ & $0.0552 \pm 0.0017$& $0.0333 \pm 0.0013$ &  \textbf{$0.0318 \pm 0.0010$} & $0.0361 \pm 0.00134$ & $0.0324 \pm 0.0009$ & $0.0384 \pm 0.0015$ & 0.0598 $\pm$ 0.0022 \\
    IS \(\uparrow\) & $1.8737 \pm 0.0244$ & $2.7361 \pm 0.0449$ & $2.1717 \pm 0.0216$ & $2.0002 \pm 0.0287$ & $1.9634 \pm 0.0200$ & $1.9736 \pm 0.0209$ & $1.8755 \pm 0.0015$ & \textbf{$2.2641 \pm 0.0271$}\\
 \bottomrule
  \end{tabular}}
	\caption{\small Metrics of the images generated from the MNIST dataset}
	\label{tab:tabella1}
\end{table}

Considering now the case of the KID metric, represented in Fig.~\ref{fig: kidmnist}, again the BasicQ architecture performs worse than all the other architectures. Also, the trend of the KID of the other architectures conforms to what has already been analyzed with FID. First, all quantum architectures have a higher KID and therefore worse compared to what is presented by the classical architecture. But already at the tenth epoch, the result is completely different: all quantum architectures (except again the BasicQ one) have a lower KID with an improvement of at least 10\% compared to the classical model. Halfway through training, the quantum models, except again BasicQ, still have the best performance. At the end of training, the results confirm as before: almost all quantum models are better than the classical one.

Finally, let's consider the IS metric, represented in Fig.~\ref{fig:ismnist}. In this case, the results are opposite to what has been analyzed so far. Indeed, the BasicQ in terms of IS performs better than any architecture, but also the rest of the quantum architectures have values significantly higher than the classical one at every point of the training. In summary, quantum architectures generally perform better than the classical one, assuming on average better values of the metrics. There are no notable differences between the various architectures used, except for BasicQ, which instead has significantly worse performance than the classical one on FID and KID. Therefore, even using a quantum architecture with a reduced depth, equal to 3, is sufficient to have better performance than the classical model.

For the sake of completeness, we also tested a GAN and a QGAN, both operating in the latent space. The QGAN used for comparison is inspired by the LaSt-QGAN introduced in \cite{chang2024latentstylebasedquantumgan}, realizing the quantum generator with the same ansatz depicted in Fig.~\ref{fig:rxrzrx} with an ansatz depth of 4 and measurement performed on the X observable. The number of parameters of the quantum generator is therefore 120, one-third compared to our Quantum Latent Diffusion Models, which is based on the use of three VQCs. The discriminator, on the other hand, is implemented classically based on the proposal in \cite{chang2024latentstylebasedquantumgan}, consisting of two hidden layers with 10 nodes each, resulting in a parameter count of 231, ensuring that the sum of parameters of the quantum generator and the classical discriminator is comparable to that of the our QLDM. 
The classical GAN implemented for comparison mirrors the QGAN, using an MLP as the classical generator model with a parameter count of 110, similar to the number of parameters in the quantum generator. Both models were trained for 40 epochs on latent vectors obtained from the MNIST dataset using the same autoencoder used for the QLDM. 
In particular, the results obtained from the QGAN and GAN, as reported in Table~\ref{tab:tabella1}, are worse compared to the results achieved by our QLDM and also inferior to the classical diffusion model in terms of FID and KID metrics. However, for IS, the QGAN demonstrates superior performance. The performance of the QGAN and GAN is limited by the extremely low number of generator parameters, which is indeed one-third of that of our QLDM.

Furthermore, we conducted an analysis in terms of dataset usage required to train the architecture. Fixing the total training to 40 epochs, we varied the percentage of the dataset on which the QLDM is trained. We considered percentages of the dataset as follows: first, 20\%, then 40\%, 60\%, 80\%, and finally the entire dataset.
First, let's consider the FID metric, depicted in Fig.~\ref{fig:fidmnistp}. With a very low percentage, equal to 20\%, the classical model outperforms two quantum models, while quantum models using observable X as a measurement show remarkable performance even with a very low dataset percentage. Increasing the percentage to 40\%, all considered quantum models surpass the performance of the classical model. Continuing to increase the percentages, quantum models continue to outperform the classical one, with the best performance always achieved by models using observable X as a measurement. Considering now the KID, shown in Fig.~\ref{fig:kidmnistp}, the results are rather similar to what was described earlier with FID. Since the case of 20\% percentage, quantum models show significantly lower KID. Finally, considering the IS, shown in Fig.~\ref{fig:ismnistp}, the previous considerations remain: the IS of the classical model is always lower at any dataset percentage considered. Thus, the proposed quantum models, in general, manage to perform better on the MNIST dataset even when trained with reduced dataset percentages, already presenting with 40\% of the training data metrics values that the classical model achieves only when trained on the entire dataset.

\subsection{Fashion MNIST dataset results}
\begin{figure}[!ht]
  \begin{subfigure}{0.5\textwidth}
    \includegraphics[width=\linewidth]{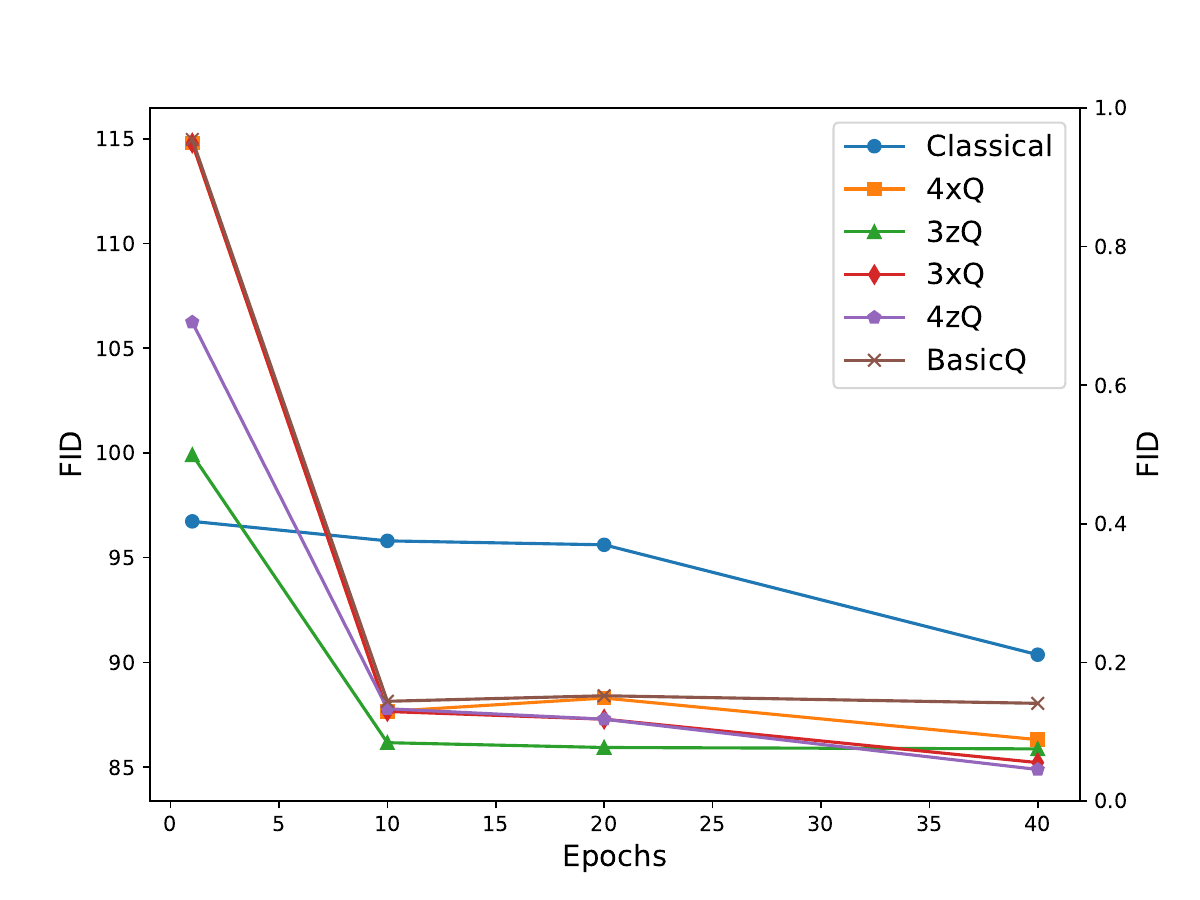}
     \caption{\small }
    \label{fig: fidfashion}
  \end{subfigure}
  \hfill
  \begin{subfigure}{0.5\textwidth}
    \includegraphics[width=\linewidth]{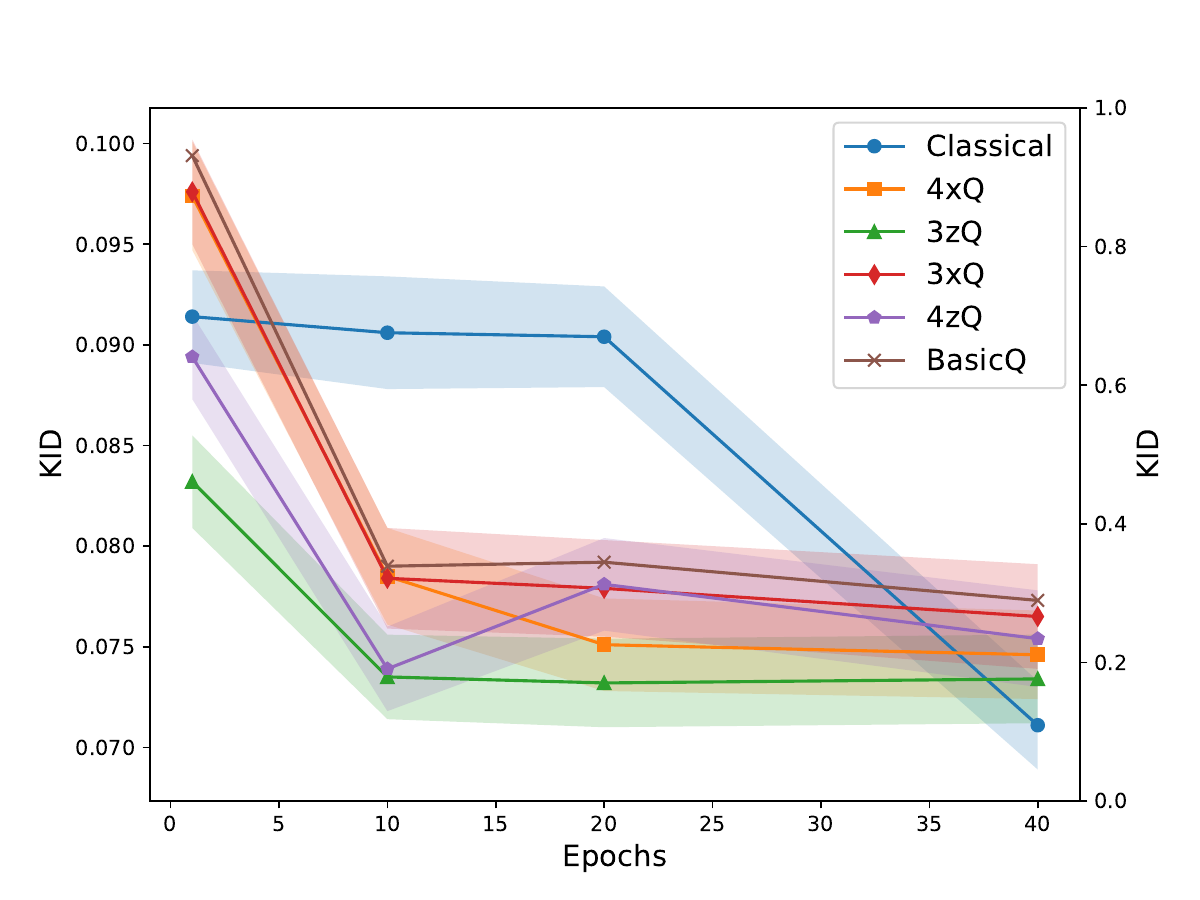}
    \caption{\small }
    \label{fig: kidfashion}
  \end{subfigure}
  \hfill
  \begin{subfigure}{0.5\textwidth}
    \includegraphics[width=\linewidth]{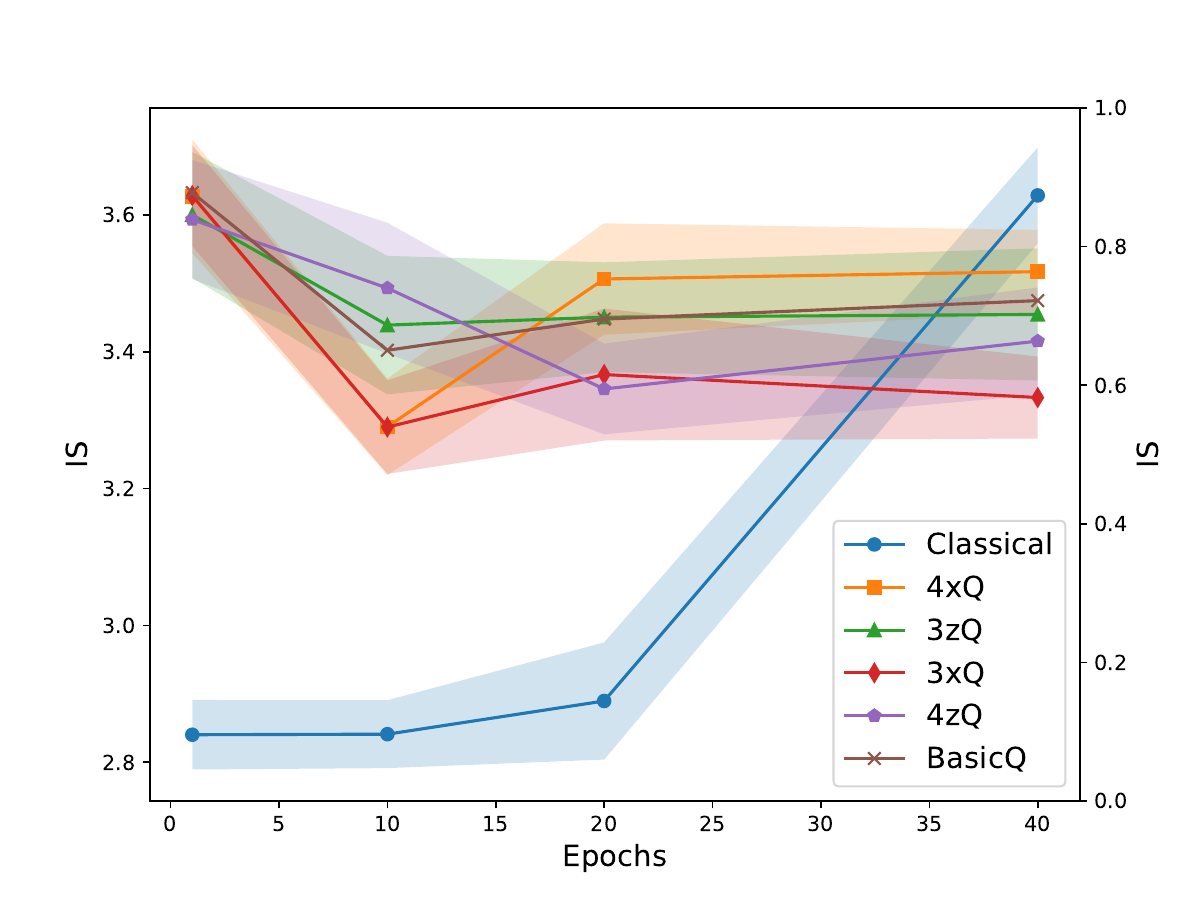}
    \caption{\small }
    \label{fig:isfashion}
  \end{subfigure}
    \hfill
  \begin{subfigure}{0.5\textwidth}
    \includegraphics[width=\linewidth]{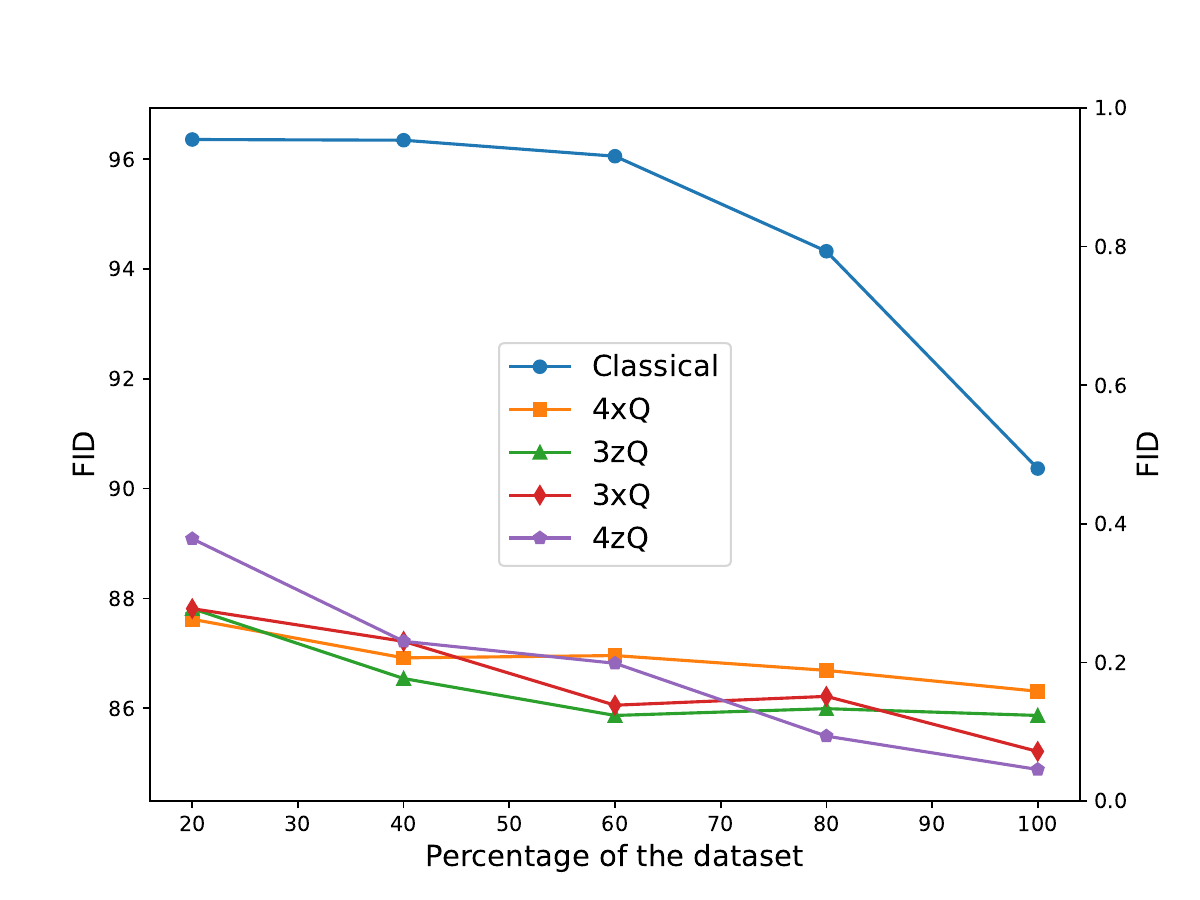}
    \caption{\small }
    \label{fig:fidfashionp}
    
  \end{subfigure}
      \hfill
  \begin{subfigure}{0.5\textwidth}
    \includegraphics[width=\linewidth]{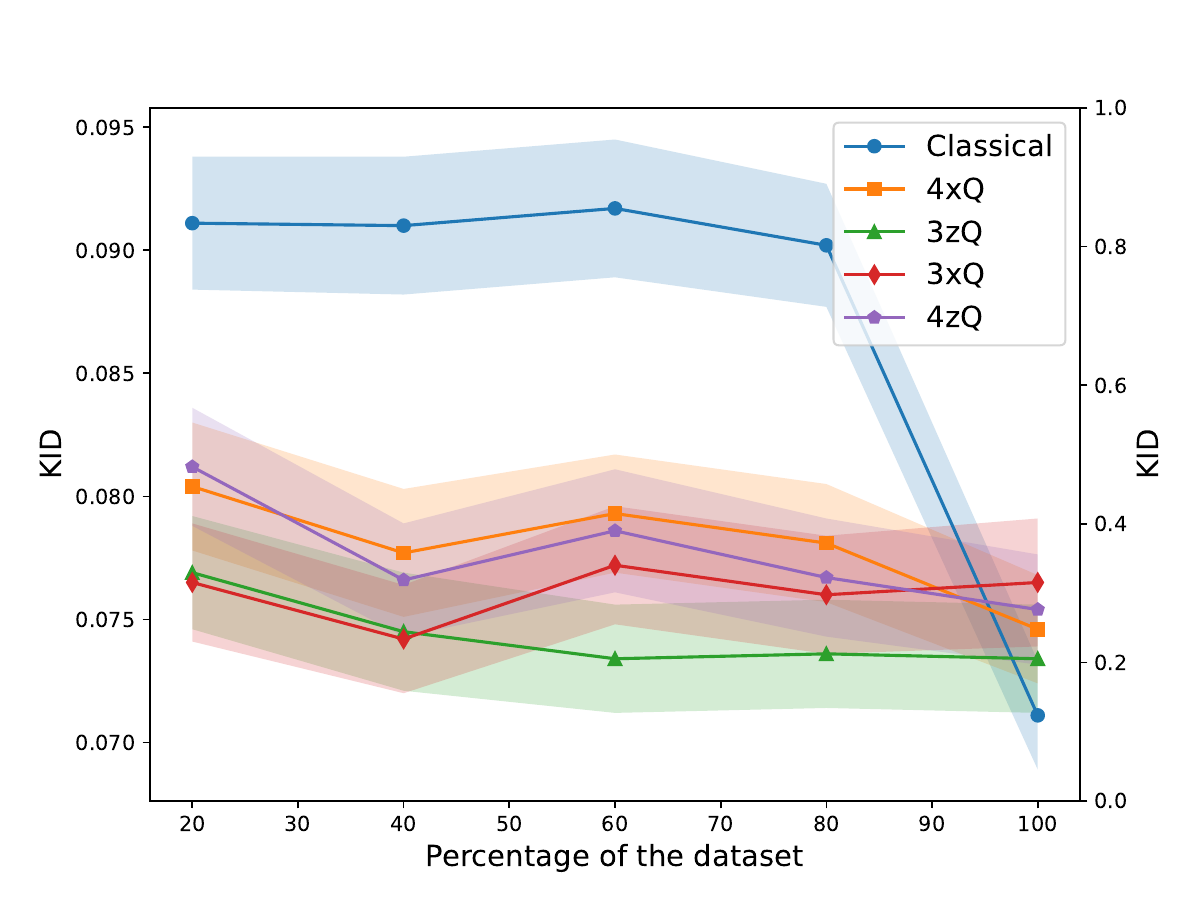}
    \caption{\small }
    \label{fig:kidfashionp}
    
  \end{subfigure}
      \hfill
  \begin{subfigure}{0.5\textwidth}
    \includegraphics[width=\linewidth]{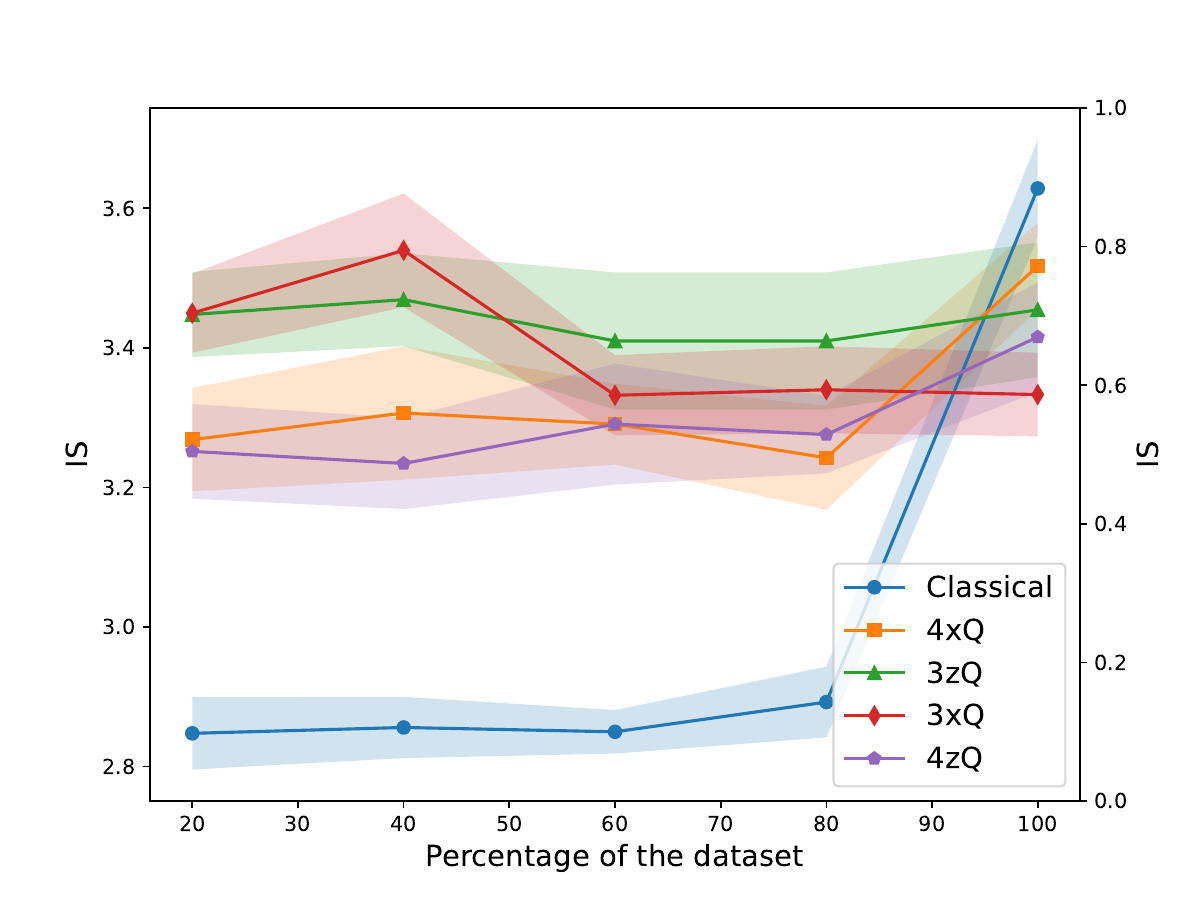}
    \caption{\small }
    \label{fig:isfashionp}
    
  \end{subfigure}
  \caption{\small Plots Fashion MNIST: a) FID plot as a function of training epochs using 100\% of the dataset for model training; b) KID plot as a function of training epochs using 100\% of the dataset for model training; c) IS plot as a function of training epochs using 100\% of the dataset for model training; d) FID plot as a function of dataset percentages; e) KID plot as a function of dataset percentages; f) IS plot as a function of dataset percentages. In the KID and IS plots, the mean and standard deviation are plotted. For the KID, 100 subsets of size 100 were used for computing mean and variance, while for IS, 10 subsets of size 1000 were used.}
  \label{fig: fashion}
\end{figure}

Let us now analyze the case where the generated images belong to the Fashion MNIST dataset. The architectures considered are the same as in the previous MNIST case. Starting again from the analysis of the generated images after the architecture has been trained for a certain number of epochs, let's first consider the FID metric, shown in Fig.~\ref{fig: fidfashion}. Again, at the first epoch, the FID value of the images generated by the classical network is better than those of the quantum models. However, at the tenth epoch, the situation is completely opposite: all the quantum models considered perform better than the classical one. It's interesting to note how the FID values at the tenth epoch of the quantum models are already lower than that obtained by the classical model at the end of training.

\begin{figure}[!ht]
  \begin{subfigure}{0.4\textwidth}
    \includegraphics[width=\linewidth]{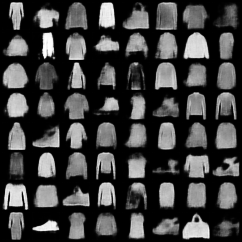}
    \caption{\small }
    \label{fig: fashion classica}
  \end{subfigure}
  \hfill
  \begin{subfigure}{0.4\textwidth}
    \includegraphics[width=\linewidth]{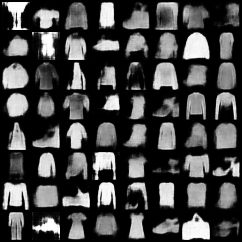}
    \caption{\small }
    \label{fig: fashion 4z}  
  \end{subfigure}
  \caption{\small Generated images of Fashion MNIST: a) generated by the Classical Architecture; b) generated by the 4zQ architecture.}
  \label{fig: fashion im}
\end{figure}

Continuing with the training, the FID of the classical model continues to decrease but does not reach the quantum models. The classical model indeed achieves an FID of 90.3655, while the 4zQ architecture achieves a lower FID of 84.8859 as shown in Table~\ref{tab:tabella2}. The images generated by the Classical Architecture and the 4zQ architecture are also shown in Fig.~\ref{fig: fashion im}.

\begin{table}[!ht]
  \centering
  \resizebox{\textwidth}{!}{%
  \begin{tabular}{lcccccc}
    \toprule
     Metrics & Classical& BasicQ & 3zQ & 3xQ & 4zQ & 4xQ\\
    \midrule
    Parameters & 330 & 120& 270 & 270& 360 &360\\
    FID \(\downarrow\) & 90.3655 & 88.0443& 85.8698 & 85.2147 & \textbf{84.8859}& 86.3102 \\
    KID \(\downarrow\) & \textbf{$0.0711 \pm 0.0022$} & $0.0773 \pm 0.0024$& $0.0734 \pm 0.0022$ &  $0.0765 \pm 0.0026$ & $0.0754 \pm 0.00224$ & $0.0746 \pm 0.0022$ \\
    IS \(\uparrow\) & \textbf{$3.6283 \pm 0.0701$} & $3.4746 \pm 0.0735$ & $3.4544 \pm 0.0966$ & $3.3330 \pm 0.0599 $&$ 3.4154 \pm 0.0784$ & $3.5169 \pm 0.0617$\\
 \bottomrule
  \end{tabular}}
	\caption{\small Metrics of the images generated from the Fashion MNIST dataset}
	\label{tab:tabella2}
\end{table}

Unlike what was observed previously in the MNIST case, the BasicQ architecture performs better than the classical model. However, due to the limited number of parameters, the FID of the BasicQ images does not improve much after the tenth epoch, and at the end of training, it is slightly worse than that of the other quantum models. 

Considering now the KID metric represented in Fig.~\ref{fig: kidfashion}, it is noticed that at the first epoch, only the models that have measurements performed with observable Z are better than the classical one. As before, at the tenth epoch, the KID values of the various quantum architectures improve significantly compared to what was obtained at the first epoch and therefore present values much lower than those obtained by the classical model.

Finally, at the end of training, it is the classical model that has the lowest and therefore best KID value, but the quantum models, even those with fewer parameters, do not deviate much. Considering the last metric, IS, whose values are represented in Fig.~\ref{fig:isfashion}, it is noted that at the first epochs, the values of the quantum models exceed by far that of the classical model. As happened for KID, at the end of training, the best IS is that of the classical model. However, even if the IS and KID of the classical model are slightly better than those of the quantum models, it is interesting to note that the quantum models take fewer epochs to reach the values they converge to. The classical model indeed requires all 40 epochs, and the variation between the performances shown at the 20th epoch and the 40th is significant. The quantum models, on the other hand, already at the tenth epoch, present excellent values on all metrics, which in the case of FID are even better than what the classical model achieves at the end of training.

We trained the models on different percentages of the dataset, as in the case of the MNIST dataset, taking initially percentages of 20\%, then 40\%, 60\%, 80\%, and finally the entire dataset. Firstly, considering the FID, whose trends are shown in Fig.~\ref{fig:fidfashionp}, it is observed that quantum models achieve better results compared to the classical model. Even with just 20\% of the dataset, the FID obtained by the quantum models significantly surpasses that obtained by the classical model trained with the entire dataset. It is interesting to note that the classical model only improves starting from 60\% of the dataset with a clear improvement when the model is trained on the entire dataset, unlike the quantum models which instead achieve excellent results already with small percentages of the dataset.

Considering instead the KID metric, represented in Fig.~\ref{fig:kidfashionp}, it is noticed that for low percentages of the dataset, the best results are obtained by all quantum models, surpassed by the classical model only when the entire dataset is considered. Similarly, for the last metric considered, IS shown in Fig.~\ref{fig:isfashionp}, the quantum models outperform the classical one for low percentages of the dataset, while when considering the entire dataset, the classical model performs better.

\subsection{EuroSAT dataset results}
\begin{figure}[!ht]
  \begin{subfigure}{0.5\textwidth}
    \includegraphics[width=\linewidth]{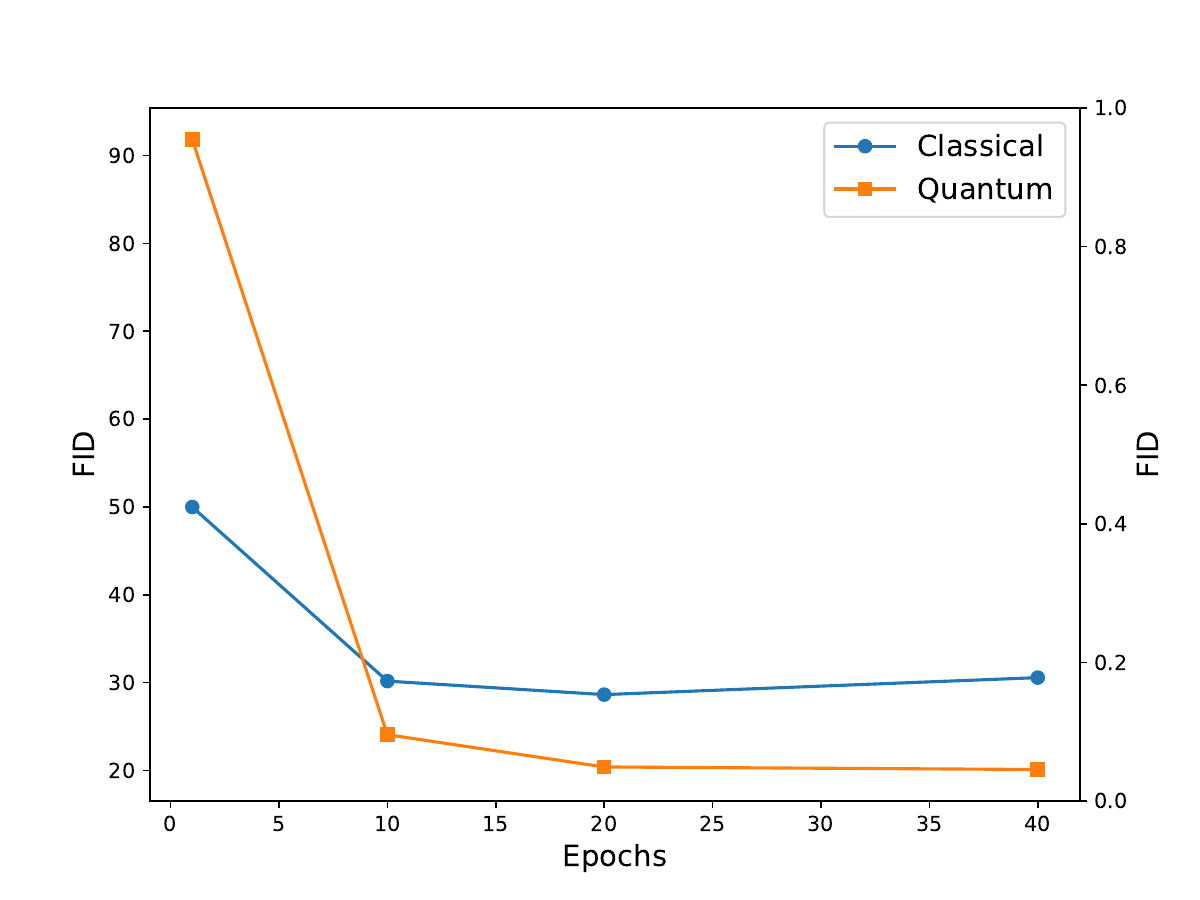}
    \caption{\small }
    \label{fig: fideurosat}
  \end{subfigure}
  \hfill
  \begin{subfigure}{0.5\textwidth}
    \includegraphics[width=\linewidth]{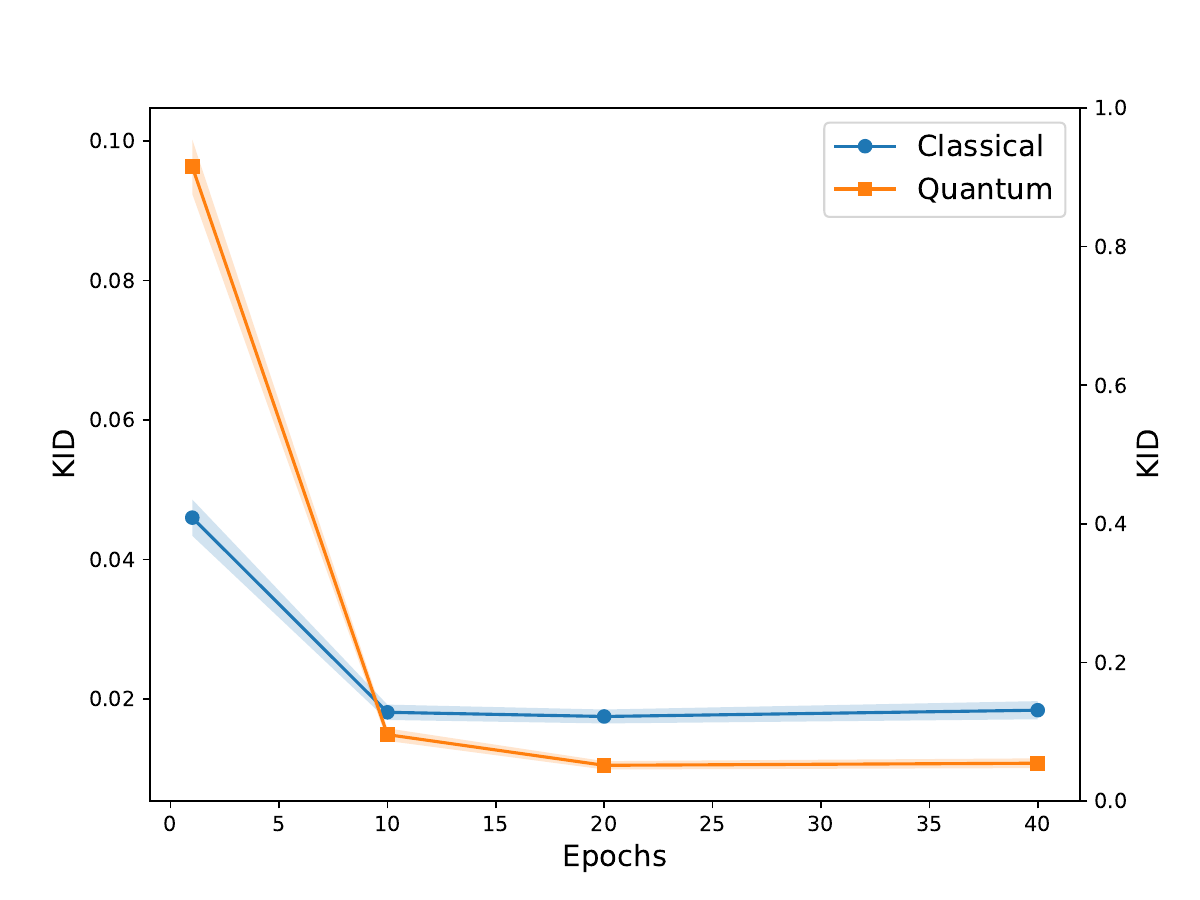}
    \caption{\small }
    \label{fig: kideurosat}
  \end{subfigure}

  \centering
  \begin{subfigure}{0.5\textwidth}
    \includegraphics[width=\linewidth]{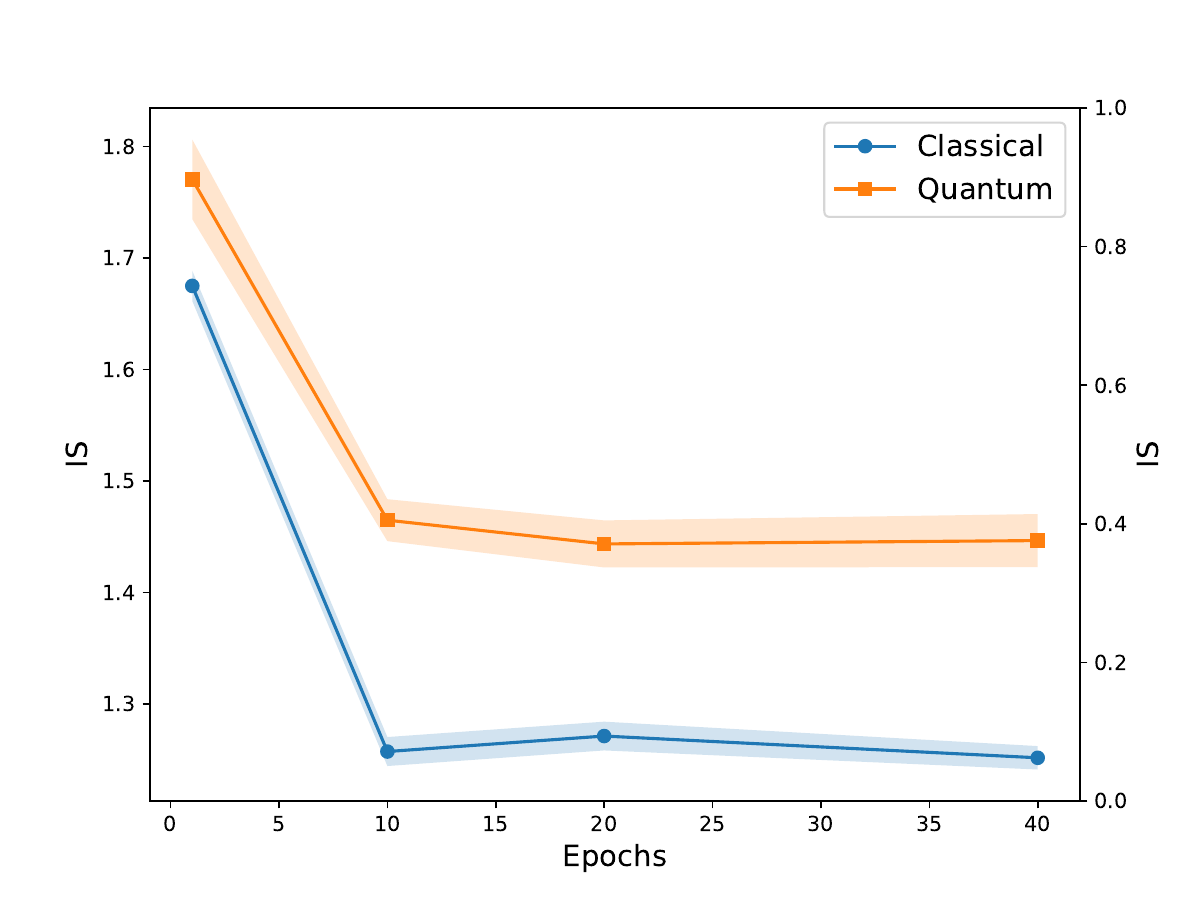}
    \caption{\small }
    \label{fig:iseurosat}
  \end{subfigure}
  
  \caption{\small Plots EuroSAT: a) FID plot as a function of training epochs using 100\% of the dataset for model training; b) KID plot as a function of training epochs using 100\% of the dataset for model training; c) IS plot as a function of training epochs using 100\% of the dataset for model training. In the KID and IS plots, the mean and standard deviation are plotted. For the KID, 100 subsets of size 100 were used for computing mean and variance, while for IS, 10 subsets of size 1000 were used.}
  \label{fig: eurosat}
\end{figure}

\begin{figure}[!ht]
  \begin{subfigure}{0.4\textwidth}
    \includegraphics[width=\linewidth]{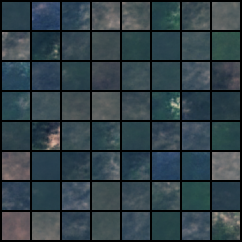}
    \caption{\small }
    \label{fig: eurosat classica}
  \end{subfigure}
  \hfill
  \begin{subfigure}{0.4\textwidth}
    \includegraphics[width=\linewidth]{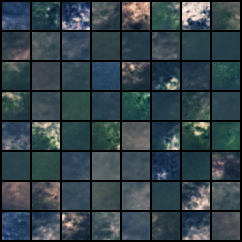}
    \caption{\small }
    \label{fig: eurosat 4z}  
  \end{subfigure}
  \caption{\small Generated images of EuroSAT: a) generated by the Classical Architecture; b) generated by the 4zQ architecture.}
  \label{fig: eurosat im}
\end{figure}

Tests were also conducted on the EuroSAT dataset to evaluate the performance of our model on real-world data, which presents different challenges compared to MNIST and Fashion MNIST. However, our model has an extremely low number of parameters, limiting its ability to handle a more complex dataset. For this reason, only two classes from the EuroSAT dataset were considered, specifically the Forest and Herbaceous Vegetation classes, with images rescaled to 28x28.
By testing the 4zQ architecture and comparing it with the classical architecture, as shown in Fig.~\ref{fig: fideurosat}, Fig.~\ref{fig: kideurosat}, and Fig.~\ref{fig:iseurosat}, it can be seen that despite the extremely low number of parameters, our model still achieves good results on a dataset containing real-world images. In particular, as visible in Fig.~\ref{fig: fideurosat}, the images generated by our quantum model initially have a higher FID, but then improve significantly in the following epochs. As noted earlier, by the tenth epoch, the images generated by the quantum model have a better FID than the images generated by the classical model after more epochs of training. The same is observed for the KID metric, as shown in Fig.~\ref{fig: kideurosat}. The last metric considered, IS, also highlights the quality of the images generated by our quantum model as shown in Fig.~\ref{fig:iseurosat}. 
The images generated at the end of training by the classical model and the 4zQ architecture are shown in Fig.~\ref{fig: eurosat im} and as shown in Table~\ref{tab:tabella_EuroSAT}, the improvements brought by using the quantum model in terms of metrics are considerable: the FID improves by more than 34\% compared to the classical model, the KID by more than 41\%, and the IS by more than 15\%.
\begin{table}[!ht]
  \centering
  \resizebox{0.5\textwidth}{!}{%
  \begin{tabular}{lcc}
    \toprule
     Metrics & Classical& 4zQ \\
    \midrule
    Parameters& 330 & 360\\ 
    FID\(\downarrow\) & 30.5643 & \textbf{20.1073}\\
    KID\(\downarrow\) &  $0.0184 \pm 0.0013$ & \textbf{$0.0108 \pm 0.0007$}\\
    IS\(\uparrow\) & $1.2517 \pm 0.0105$ & \textbf{$1.4465 \pm 0.0238$}\\
 \bottomrule
  \end{tabular}}
  \caption{\small Metrics of the images generated from the EuroSAT dataset}
  \label{tab:tabella_EuroSAT}
\end{table}

\subsection{Hyperparameter analysis}
We also conducted some experiments to explore the QLDM's sensitivity to certain hyperparameters. As we have already seen in previous tests where different architectures consisting of 3 or 4 layers were tested on the MNIST and Fashion MNIST datasets, the number of layers does not seem to significantly affect performance. While increasing the depth from 3 to 4 layers provided a slight performance improvement, the gains were marginal. This suggests that a deeper circuit can enhance expressivity, but the benefits plateau beyond a certain depth, likely due to computational complexity. 
Similarly, the performance of our QLDM was comparable across these different measurement bases, indicating robustness in the choice of measurement gates.

\begin{figure}[!ht]
  \begin{subfigure}{0.5\textwidth}
    \includegraphics[width=\linewidth]{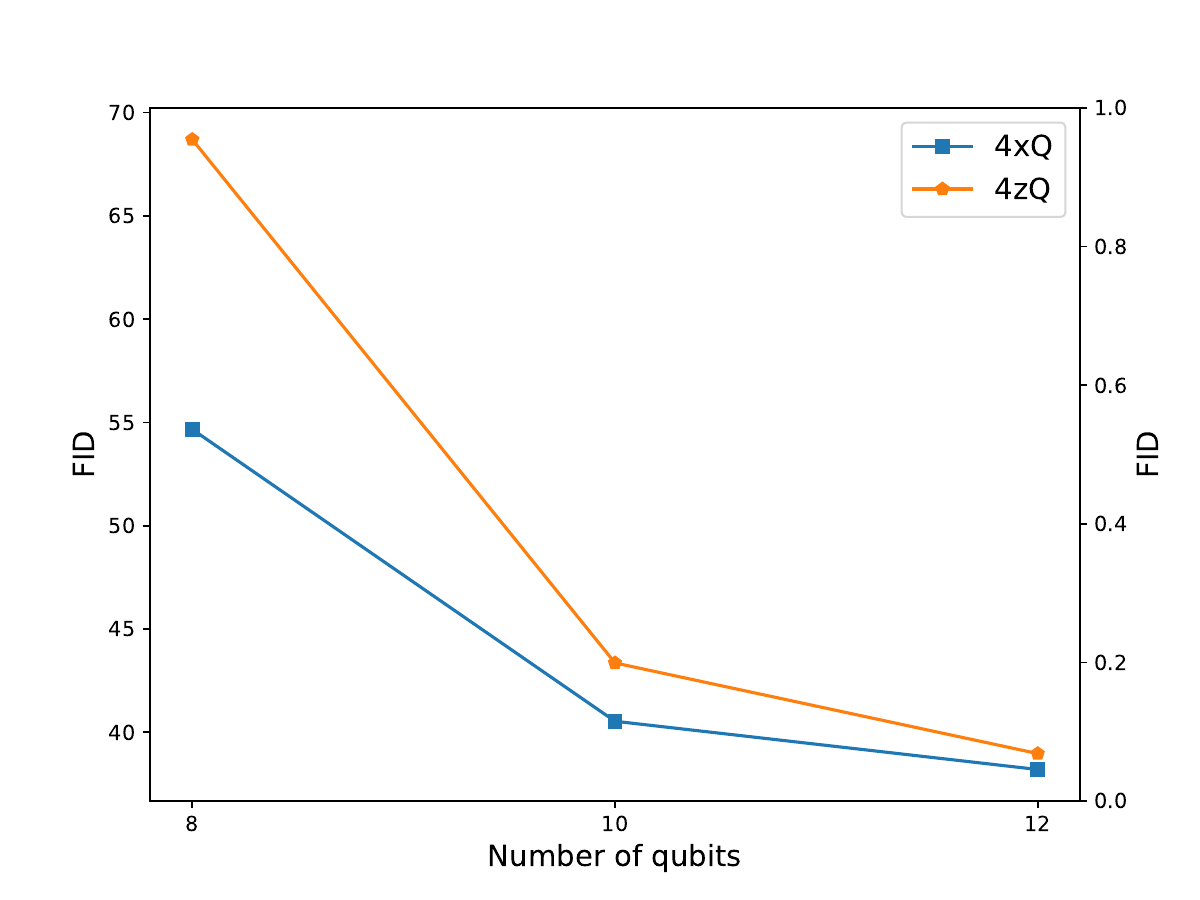}
    \caption{\small }
    \label{fig: fidqubits}
  \end{subfigure}
  \hfill
  \begin{subfigure}{0.5\textwidth}
    \includegraphics[width=\linewidth]{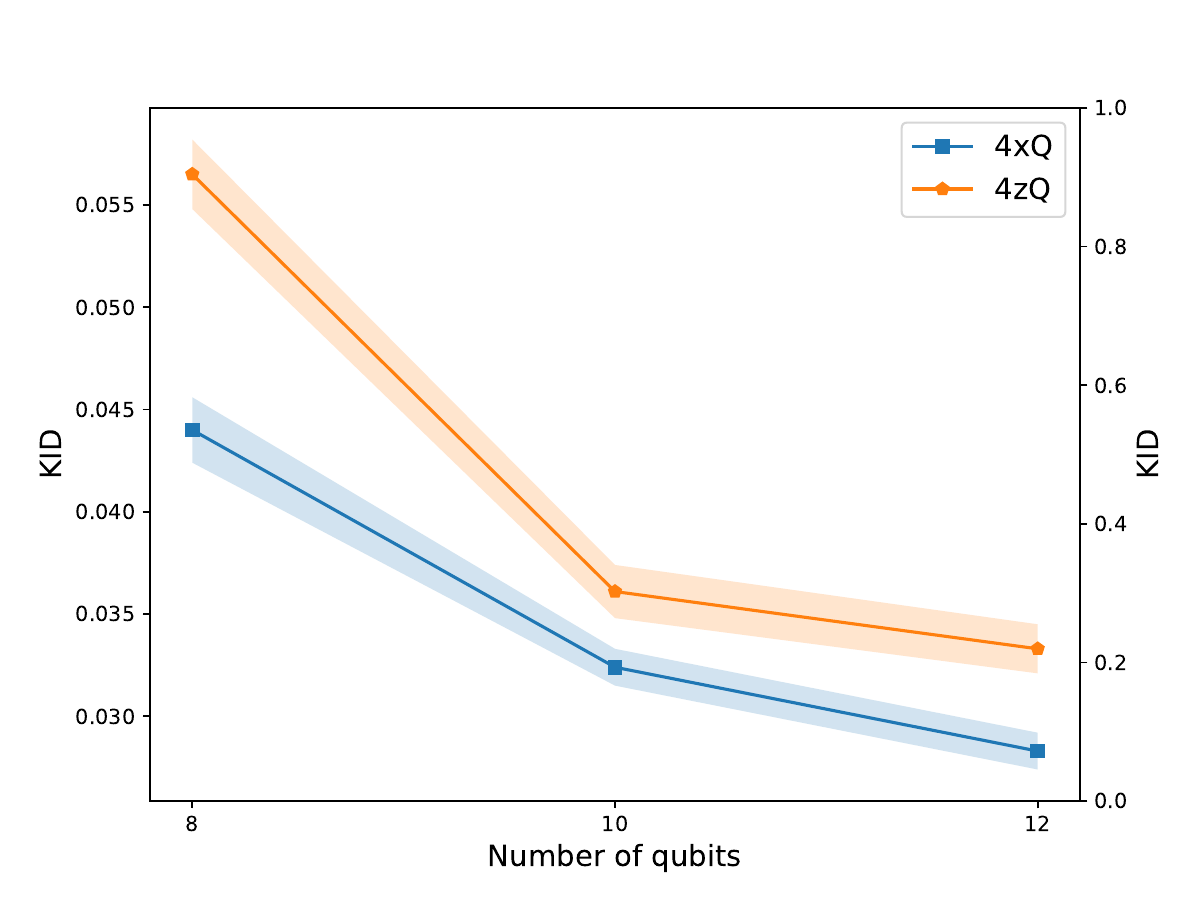}
    \caption{}
    \label{fig: kidqubits}
  \end{subfigure}

  \centering
  \begin{subfigure}{0.5\textwidth}
    \includegraphics[width=\linewidth]{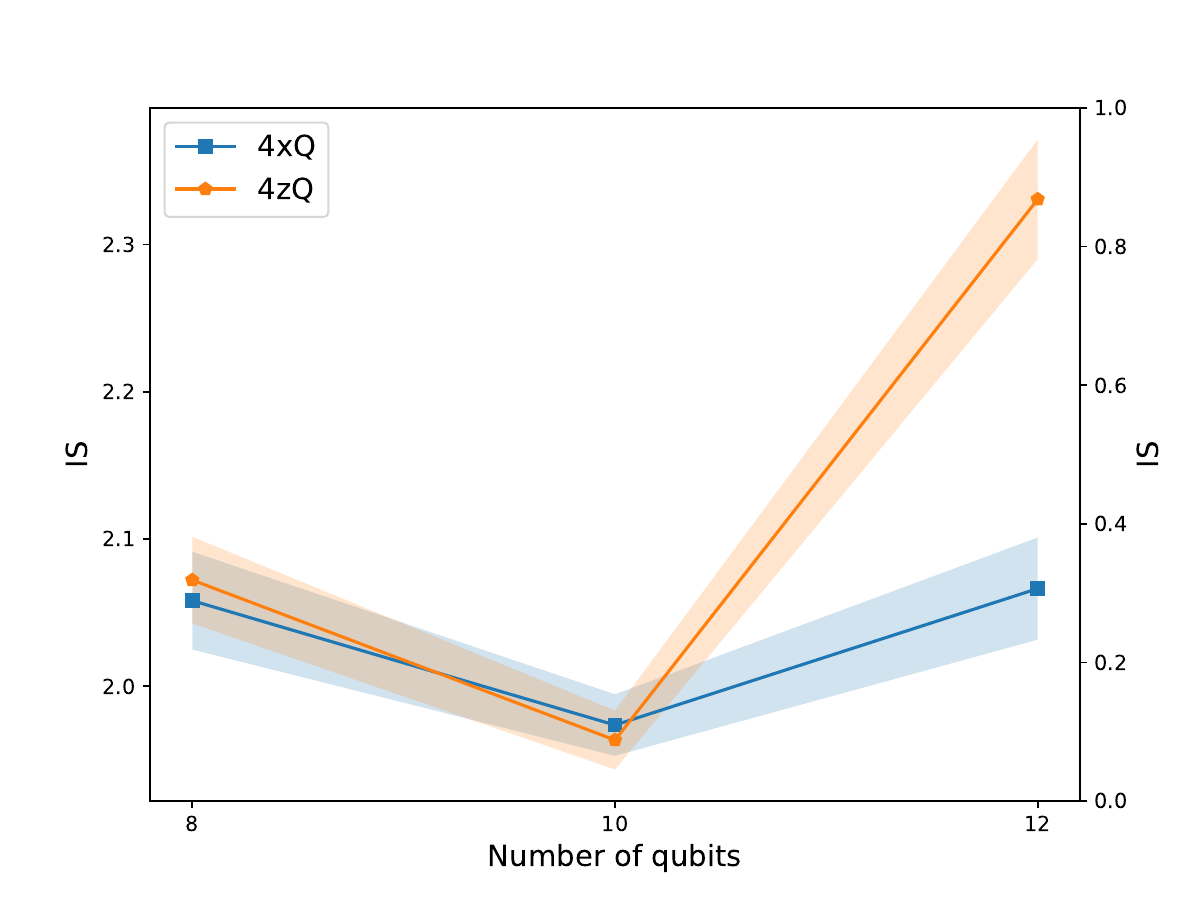}
    \caption{\small }
    \label{fig:isqubits}
  \end{subfigure}
  
  \caption{\small Plots showing the sensitivity of our model to variations in the number of qubits:
a) Trends of the FID metric computed on the generated images from the MNIST dataset by the architectures 4xQ (in blue) and 4zQ (in orange) after training for 40 epochs on 100\% of the dataset, considering different numbers of qubits, namely 8, 10, and 12; b) Trends of the KID metric computed on the generated images from the MNIST dataset by the architectures 4xQ (in blue) and 4zQ (in orange) after training for 40 epochs on 100\% of the dataset, considering different numbers of qubits, namely 8, 10, and 12; c) Trends of the IS metric computed on the generated images from the MNIST dataset by the architectures 4xQ (in blue) and 4zQ (in orange) after training for 40 epochs on 100\% of the dataset, considering different numbers of qubits, namely 8, 10, and 12.}
  \label{fig: qubits}
\end{figure}

The number of qubits, on the other hand, proves to be more influential. By testing the 4zQ and 4xQ architectures on the MNIST dataset with three different numbers of qubits, namely 8, 10, and 12, it is evident from Table~\ref{tab:tabella_MNIST_qubits} that, as the number of qubits increases, the performance improves. This is due to the more expressive latent space representation. In particular, as evidenced by the analysis of the graphs shown in Fig.~\ref{fig: fidqubits}, Fig.~\ref{fig: kidqubits}, and Fig.~\ref{fig:isqubits}, which respectively show the FID, KID, and IS metrics of the images generated by the 4xQ and 4zQ architectures after being trained for 40 epochs, there is a significant improvement in performance when moving from 8 to 10 qubits, and a smaller yet present improvement when moving from 10 to 12 qubits. However, increasing the number of qubits significantly increases the required computational cost. For this reason, balancing computational cost and performance obtained, we chose to maintain the number of qubits at 10 in the tests conducted in the previous sections. 

\begin{table}[!ht]
  \centering
  \resizebox{0.8\textwidth}{!}{%
  \begin{tabular}{lcccc}
    \toprule
     Architecture & number of qubits & FID \(\downarrow\) & KID\(\downarrow\) & IS \(\uparrow\)\\
     \midrule
     4zQ & 8 & 68.6946 & $0.0565 \pm 0.0017$ & $2.0721 \pm 0.0294$\\ 

     4xQ & 8 & 54.6635 & $0.0440 \pm 0.0016$ & $2.0581 \pm 0.0332$ \\
  
     4zQ & 10 &  43.3555 & $0.0361 \pm 0.0013$ & $1.9634 \pm 0.0200$ \\
  
     4xQ & 10 & 40.5321 & $0.0324 \pm 0.0009$ & $1.9736 \pm 0.0209$\\
  
     4zQ & 12 & \textbf{38.9669} & \textbf{$0.0333 \pm 0.0012$} & \textbf{$2.3308 \pm 0.0406$} \\
  
     4xQ & 12 & \textbf{38.1971} & \textbf{$0.0283 \pm 0.0009$}& \textbf{$2.0663 \pm 0.0347$}\\
 \bottomrule
  \end{tabular}}
	\caption{\small Metrics of the images generated from the MNIST dataset considering different numbers of qubits.}
	\label{tab:tabella_MNIST_qubits}
\end{table}

\begin{figure}[!ht]
  \begin{subfigure}{0.5\textwidth}
    \includegraphics[width=\linewidth]{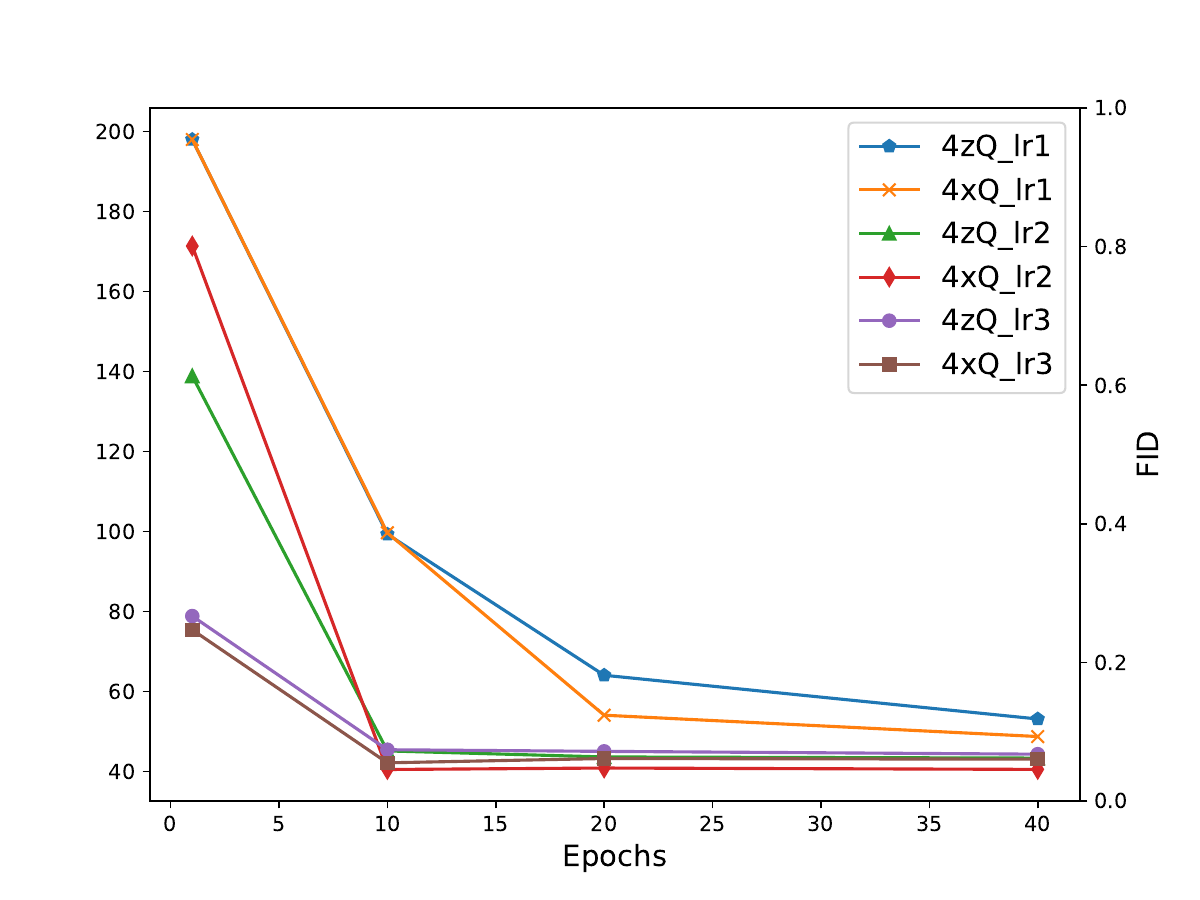}
    \caption{\small }
    \label{fig: fidlr}
  \end{subfigure}
  \hfill
  \begin{subfigure}{0.5\textwidth}
    \includegraphics[width=\linewidth]{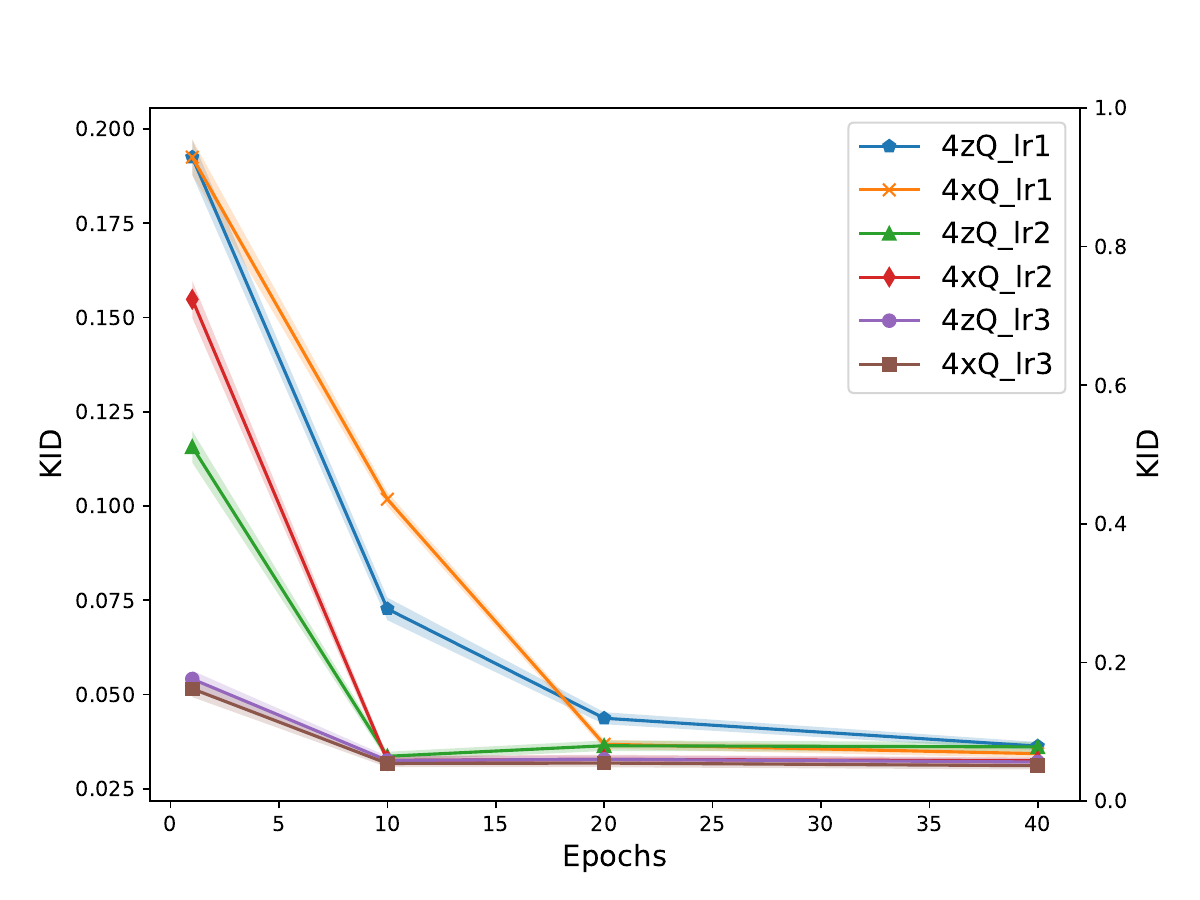}
    \caption{\small }
    \label{fig: kidlr}
  \end{subfigure}

  \centering
  \begin{subfigure}{0.5\textwidth}
    \includegraphics[width=\linewidth]{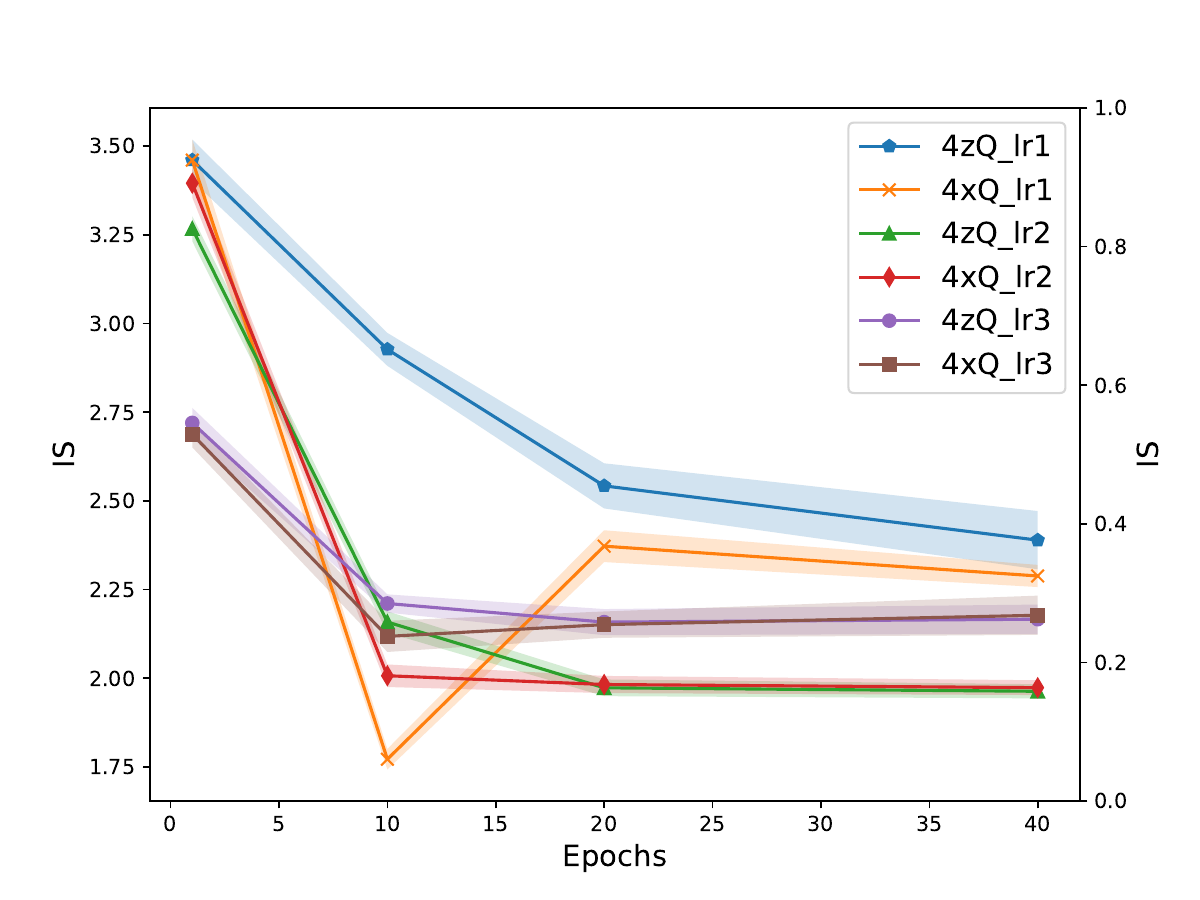}
    \caption{\small }
    \label{fig:islr}
  \end{subfigure}
  
  \caption{\small Graphs showing the sensitivity of our model to variations in learning rate. Three different learning rates were tested: lr1 where the learning rate is \(2 \times 10^{-4}\), lr2 where the learning rate is \(10^{-3}\), and lr3 where the learning rate is \(5 \times 10^{-3}\). The architectures 4zQ and 4xQ were evaluated on the MNIST dataset and trained for 40 epochs using 100\% of the dataset. Specifically:
a) Trend of the FID metric for the two different architectures with the three different learning rates;
b) Trend of the KID metric for the two different architectures with the three different learning rates;
c)Trend of the IS metric for the two different architectures with the three different learning rates.}
  \label{fig: lr}
\end{figure}

Additionally, we conducted further tests to evaluate the sensitivity of our model to the learning rate. Starting from the learning rate $\mathrm{lr2}=10^{-3}$ used in previous tests, we tested both 4zQ and 4xQ architectures on MNIST using two further learning rates; the former one ${\mathrm{lr1}=2 \cdot 10^{-4}}$ was set to one-fifth of lr2, while the other learning rate ${\mathrm{lr3}=5 \cdot 10^{-3}}$ was chosen for the sake of comparison five times lr2.
The analysis of metrics computed on images generated by the two architectures with different learning rates, as shown in Fig.~\ref{fig: fidlr}, Fig.~\ref{fig: kidlr} and Fig.~\ref{fig:islr}, reveals that using a higher learning rate (lr3) results in better images at the first epoch, but the model shows less improvement compared to models trained with the default setup (lr2). On the other hand, using a lower learning rate (lr1) results in significantly worse FID and KID metrics in the first epoch and generally poorer performance across all epochs. Ultimately, as evidenced by the results reported in Table~\ref{tab:tabella4}, which presents the metric values of images generated by models trained for 40 epochs, by using the lr2 learning rate a better performance is achieved, particularly in terms of FID.

\begin{table}[!ht]
  \centering
	\resizebox{0.8\textwidth}{!}{%
  \begin{tabular}{lcccc}
    \toprule
     Architecture & learning rate & FID \(\downarrow\) & KID\(\downarrow\) & IS \(\uparrow\)\\
     \midrule
     4zQ & $2 \cdot 10^{-4}$ & 53.1363 & $0.0363 \pm 0.0011$ & \textbf{$2.3892 \pm 0.0821$}\\

     4xQ & $2 \cdot 10^{-4}$ & 48.7490 & $0.0343 \pm 0.0011$ & \textbf{$2.2883 \pm 0.0312$}\\
  
     4zQ & $10^{-3}$ &  \textbf{43.3555} & $0.0361 \pm 0.0013$ & $1.9634 \pm 0.0200$ \\
  
     4xQ & $10^{-3}$ & \textbf{40.5321} & $0.0324 \pm 0.0009$ & $1.9736 \pm 0.0209$\\
  
     4zQ & $5 \cdot 10^{-3}$ & 44.3328 & \textbf{$0.0321 \pm 0.0010$} & $2.1665 \pm 0.0412$ \\
  
     4xQ & $ 5 \cdot 10^{-3}$ & 43.0839 & \textbf{$0.0311 \pm 0.0010$} & $2.1776 \pm 0.0553$\\
 \bottomrule
  \end{tabular}}
\caption{\small Metrics of the images generated from the MNIST dataset considering various learning rates.}
\label{tab:tabella4}
\end{table}

\subsection{Convergence analysis}
So far, an analysis has been conducted on the improvements brought by quantum architectures by considering only the quality of the images generated by quantum architectures compared to the classical one. We now conduct a further analysis paying attention to the loss obtained during the training of the models. 

\begin{figure}[!ht]
    \centering
  \begin{subfigure}{0.8\textwidth}
    \includegraphics[width=\linewidth]{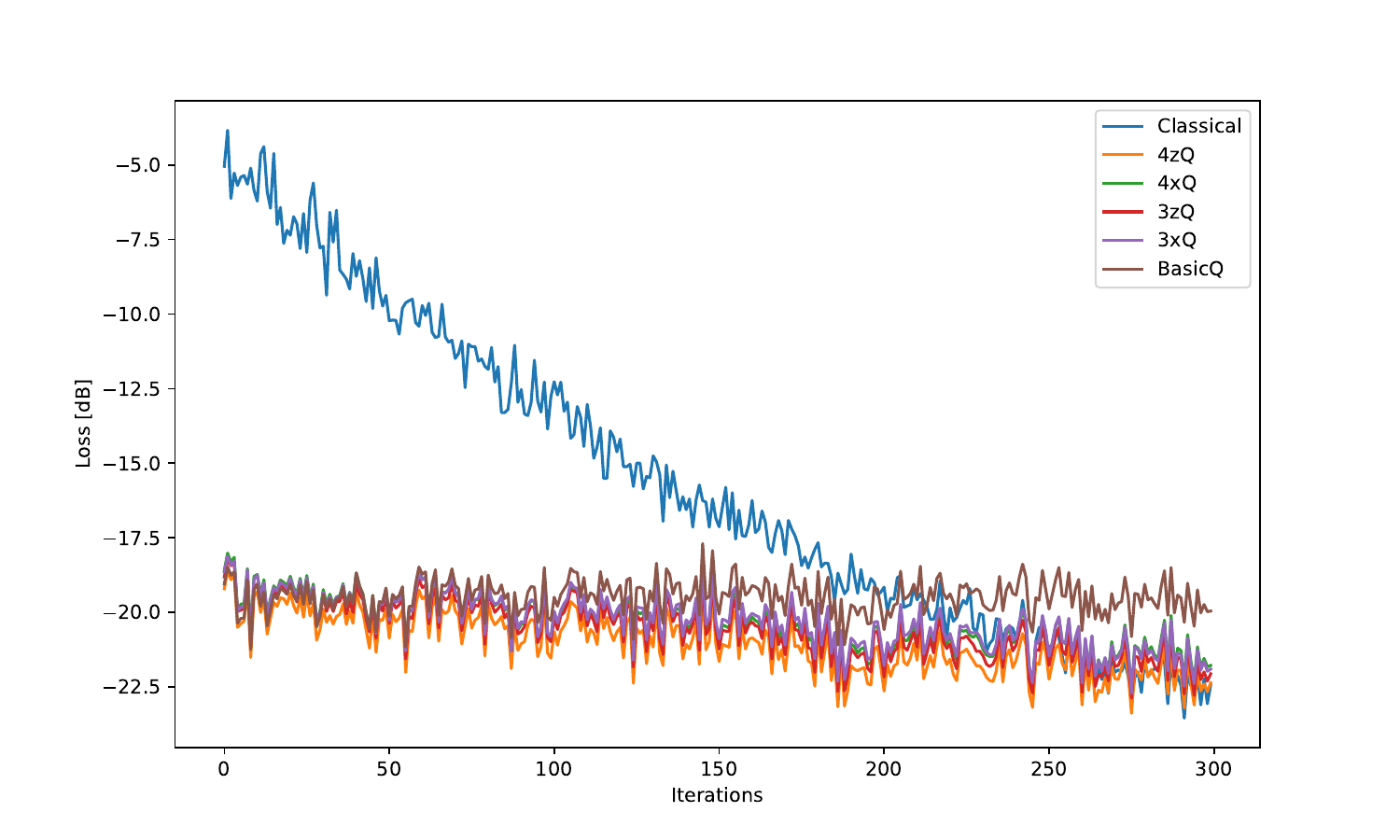}
    \caption{\small  }
    \label{fig:subfig2A}
  \end{subfigure}
  \hfill
 
  \begin{subfigure}{0.8\textwidth}
    \includegraphics[width=\linewidth]{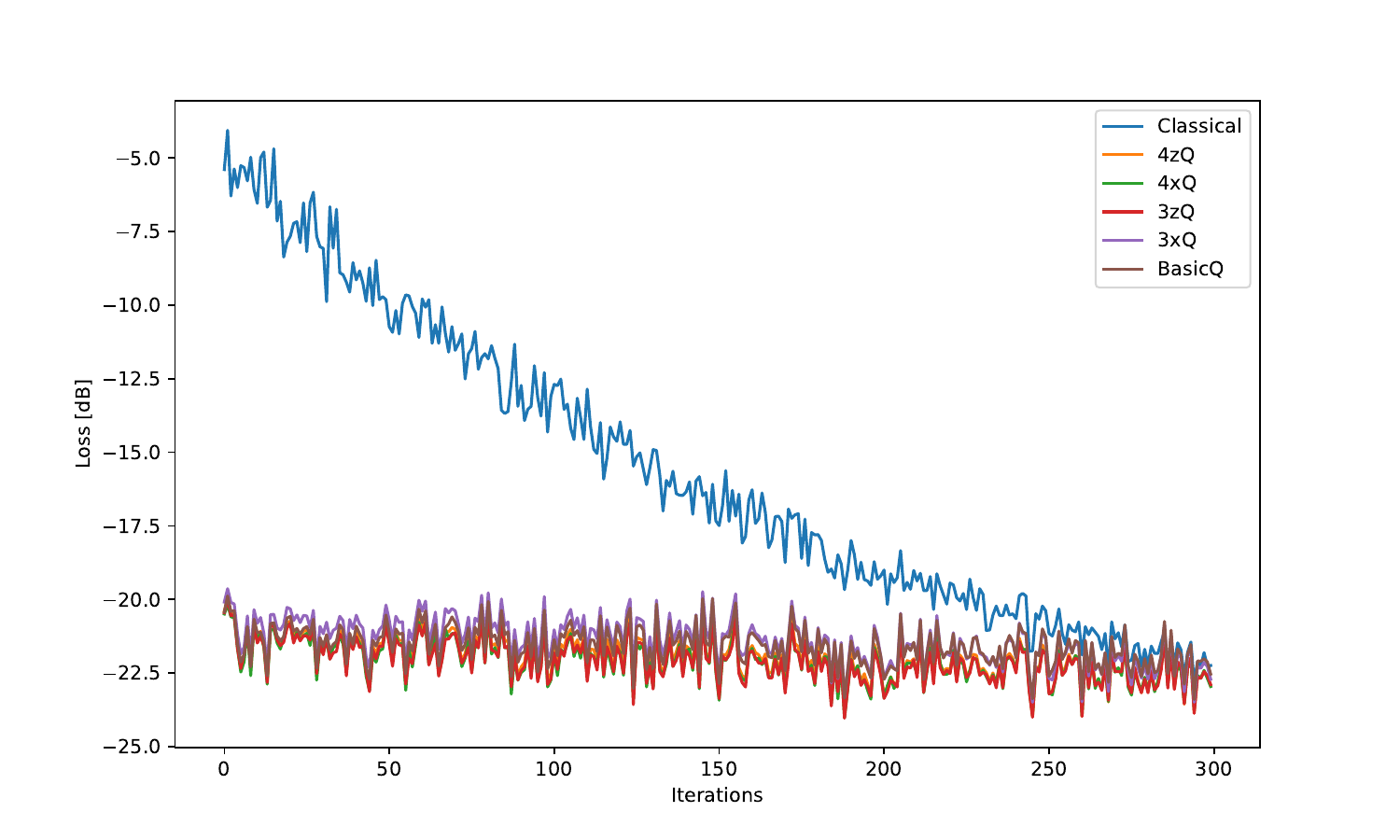}
    \caption{\small  }
    \label{fig:subfig2B}
  \end{subfigure}
   \begin{subfigure}{0.8\textwidth}
    \includegraphics[width=\linewidth]{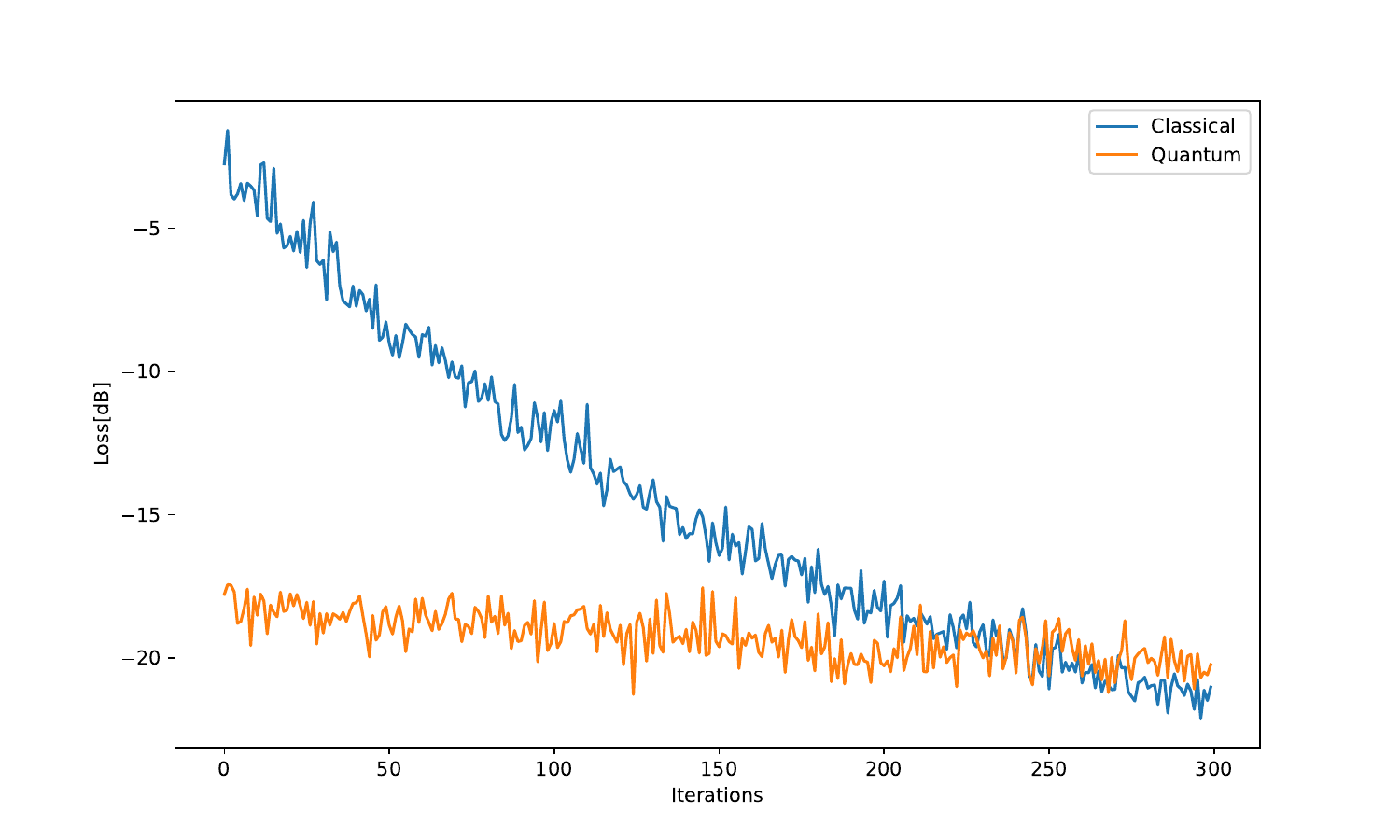}
    \caption{\small  }
    \label{fig:subfig2C}
  \end{subfigure}
    \caption{\small Loss  (in dB) of the latent diffusion model: a) details of the initial iterations for MNIST; b) details of the initial iterations for Fashion MNIST; c) details of the initial iterations for EuroSAT dataset.}
    \label{loss}
\end{figure}

As evident from the analysis of the graphs shown in Fig.~\ref{loss}, quantum architectures, both those using the basic ansatz shown in Fig.~\ref{fig: basic ansatz} and those using the ansatz shown in Fig.~\ref{fig:rxrzrx}, have from the very first iterations an extremely lower loss value, approximately an order of magnitude less, compared to the loss shown by the classical architecture. The loss shown by the classical architecture is initially very high and begins to decrease, reaching, however, after several iterations, the values obtained by the two quantum architectures. The loss of the latter, therefore, converges much earlier. This continues to be evident, as seen in Fig.~\ref{fig:subfig2C}, even when considering a more complex dataset such as EuroSAT.
This result is extremely important because it demonstrates that quantum architectures not only achieve better performance in terms of image quality, but also that the advantages of quantum architectures lie in the shorter convergence times of the loss.

\FloatBarrier

\subsection{Impact of noise}
While our current study assumes a noiseless quantum computing environment due to practical constraints, we recognize the significance of understanding and addressing noise in real-world applications. Here we explore the potential impacts of noise on the performance of our QLDM and propose preliminary strategies for mitigating these effects, setting the stage for future empirical studies. Quantum noise can arise from various sources, including decoherence, gate errors, and measurement errors \cite{resch2021benchmarking}. These types of noise can adversely affect the fidelity and robustness of quantum computations, which are critical for the effective functioning of QLDMs.

In the context of QLDMs, decoherence can reduce the model's ability to run long circuits and capture complex data correlations, thereby impacting the quality of the generated samples. Gate errors, on the other hand, can cause deviations from the intended operations. They accumulate as the depth of the quantum circuit increases, leading to significant deviations from the desired quantum state transformations. For QLDMs, this means that the transformations in the latent space may not be accurately performed, resulting in degraded model performance. Finally, inaccuracies during the readout process can lead to incorrect measurement outcomes, which can in turn affect the accuracy of the generated data.

Conversely to existing quantum diffusion models in the literature \cite{cacioppo2023quantum,parigi2023quantumnoisedriven}, our approach employs three distinct VQCs for processing latent vectors, temporal embeddings, and their combination. This architecture results in shorter circuit lengths compared to models utilizing a single VQC for all tasks or for in-circuit denoising. Moreover, our VQCs are designed based on hardware-efficient ansatzes, with linear or circular entanglement among qubits, which are well-suited for the topological constraints of current quantum devices. Each VQC in our model uses $n \times l$ parameters for the basic ansatz, where $l$ represents the number of layers, and $3n \times l$ parameters for the more expressive ansatz, providing substantial expressive power while maintaining manageability in circuit depth. Specifically, the circuit is restricted to 3 layers for the basic ansatz and 4 layers for the more complex ansatz. Such a limited depth ensures that our model remains feasible for implementation on NISQ devices according to recent literature and hardware capabilities \cite{guo2022quantum,gujju2024quantum}, with a total cumulative circuit depth of $1 + (1 + n) \times l$ for the basic ansatz circuit and of $1 + 3n \times l$ for the more complex one, respectively.
 
Beyond the design specifics of our architecture, several strategies could be adopted to mitigate the issues posed by quantum noise. The implementation of Quantum Error Correction (QEC) schemes, such as the surface code, helps in detecting and correcting errors during computation \cite{fowler2012surface}, guaranteeing reliable logical qubits. While QEC schemes are resource-intensive, they are crucial for long-term sustainability of quantum computations, as pointed out in \cite{google2023suppressing}. Additionally, noise mitigation techniques like zero-noise extrapolation, probabilistic error cancellation and clifford data regression could also be explored \cite{wang2024can}. While these approaches do not require the full overhead of QEC, they are proven to significantly reduce the impact of noise on VQCs' performance. Notably, there is numerical evidence suggesting that these techniques not only mitigate noise but may also facilitate the training process in scenarios where the cost concentration is not too severe \cite{cai2023quantum}. 
Overall, while our initial results are promising, a detailed examination of noise underscores the critical need for empirical testing and discussion of noise mitigation strategies for our QLDM. As quantum technology progresses, our future work will focus on conducting thorough experiments to evaluate the impact of noise on QLDM and improving noise mitigation techniques tailored for our use case.

\section{Conclusions}\label{sec:conclusions}
The introduction of quantum computing into generative machine learning models can bring numerous advantages, such as reducing the number of epochs required for learning, reducing the number of trainable parameters, or reducing the dataset needed to train the model. In this paper, we propose an efficient use of quantum computing within diffusion models, presenting our QLDM. Our idea is to incorporate VQCs as efficiently as possible, by first leveraging a classical convolutional autoencoder to transition from pixel space to latent space. This not only allows us to implement our QLDM with a limited number of qubits, enabling the adoption of angle encoding as data encoding, but also introduces a richer non-linearity that our quantum model can benefit from.

During testing, we analyzed different ansatz with varying depths as well as different measurement observables, always comparing them with a classical model. The results obtained on MNIST immediately demonstrate how quantum models can achieve better performance than the classical one, obtaining better metric values on almost all tested quantum architectures. Furthermore, by analyzing how performance varies with different percentages of the dataset used for training, it is evident that for the IS and KID metrics, quantum models already outperform the classical model with low dataset percentages. Additionally, quantum models achieve excellent performance even on FID with low percentages.
These results effectively demonstrate the importance of adopting quantum techniques within diffusion processes, as they allow us to achieve good performance even with extremely limited training datasets.

On the second dataset considered, Fashion MNIST, the results are in line with what has been observed. Quantum models all exhibit improved FID compared to the classical one, and moreover, they already outperform the classical model by the tenth epoch. This once again underscores the advantage of incorporating quantum techniques into these generative models. Even for the KID and IS metrics, despite the classical model showing better values at the end of training, a similar advantage can be observed. In fact,  at the tenth epoch quantum models reach comparable, albeit slightly worse, values to those assumed by the classical model at the end of the training.

Furthermore, by analyzing how performance varies with different percentages of the dataset, it is observed again how quantum models can achieve good performance even with limited training data. In particular, the FID obtained by quantum models trained with only 20\% of the dataset is better than that obtained by the classical model trained with the entire dataset.
Finally, even an analysis of the loss has shown that, from the early iterations, the quantum model achieves significantly lower loss values compared to its classical counterpart, converging to the final loss value earlier than the classical model.

Even on the last dataset considered (i.e., EuroSAT), despite the complexity of working with RGB images instead of just grayscale, although mitigated by using only two classes instead of the entire dataset, it is evident that our QLDM outperforms a classical model in terms of metrics. In fact, the images generated by the quantum model at the tenth epoch are already better than those generated by the classical model at the last epoch, once again demonstrating the ability of quantum models to learn faster and more effectively. This certainly paves the way for using our model not only for toy problems but also for real and more complex problems.

Possible future works starting from this study may involve the conditioning the model for improving the overall performance according to the specific application. Additionally, another goal may be also to enrich and expand our QLDM model in order to extend the analysis to more complex datasets, as well as to other learning tasks such as anomaly detection and time series analysis. Finally, we also aim at empirically investigating the impact of noise sources on our QLDM with a real quantum hardware.

\section*{Acknowledgments}

The Authors would like to express their sincere gratitude to Dr. Su Yeon Chang (CERN and EPFL - Ecole Polytechnique Federale Lausanne, Switzerland, su.chang@epfl.ch) who generously dedicated her time and expertise to provide constructive feedback on this research paper. Her insightful comments and suggestions have significantly enhanced the quality and rigor of this study.

The contribution of M. Panella, A. Ceschini and F. De Falco in this work was in part supported by the ``NATIONAL CENTRE FOR HPC, BIG DATA AND QUANTUM COMPUTING'' (CN1, Spoke 10) within the Italian ``Piano Nazionale di Ripresa e Resilienza (PNRR)'', Mission 4 Component 2 Investment 1.4 funded by the European Union - {NextGenerationEU} - CN00000013 - CUP B83C22002940006.

%% BioMed_Central_Bib_Style_v1.01

\end{document}